\documentclass[preprint,12pt]{elsarticle}

\makeatletter
\def\ps@pprintTitle{%
  \let\@oddhead\@empty
  \let\@evenhead\@empty
  \def\@oddfoot{\reset@font\hfil}%
  \let\@evenfoot\@oddfoot}
\makeatother

\usepackage{comment}
\usepackage{color}
\usepackage[colorlinks=true,breaklinks=true,bookmarks=true,urlcolor=blue,citecolor=blue,linkcolor=blue,bookmarksopen=false,draft=false]{hyperref}
\usepackage{algorithm}
\usepackage{algcompatible}
\usepackage{graphicx}

\graphicspath{{Figures/}}
\DeclareGraphicsExtensions{.pdf,.png,.jpg}
\setkeys{Gin}{draft=false}

\usepackage{caption}
\usepackage{multirow,booktabs,dcolumn}
\usepackage[export]{adjustbox}
\usepackage[utf8]{inputenc}
\usepackage{mathrsfs}
\usepackage{orcidlink}
\usepackage{pdflscape}

\DeclareMathAlphabet{\mathdutchcal}{U}{dutchcal}{m}{n}
\SetMathAlphabet{\mathdutchcal}{bold}{U}{dutchcal}{b}{n}
\DeclareMathAlphabet{\mathdutchbcal}{U}{dutchcal}{b}{n}
\usepackage{amssymb}
\usepackage{amsthm}
\usepackage{amsmath}
\usepackage{float}
\usepackage{epstopdf}
\usepackage[caption=false]{subfig}

\theoremstyle{plain}

\theoremstyle{definition}

\theoremstyle{remark}

\newcommand{\executeiffilenewer}[3]{%
	\ifnum\pdfstrcmp{\pdffilemoddate{#1}}%
	{\pdffilemoddate{#2}}>0%
	{\immediate\write18{#3}}\fi%
}

\journal{Submitted to Machine Learning: Science and Technology (IOP)}

\begin{document}

\begin{frontmatter}



\title{Differentiable Surrogate for Detector Simulation and Design with Diffusion Models}
\tnotetext[t1]{Submitted to Machine Learning: Science and Technology (IOP).}


\author[aff1,aff3]{Xuan Tung Nguyen\corref{cor1}\orcidlink{0009-0009-6527-441X}}
\cortext[cor1]{Correspondence: xuantung.nguyen@pd.infn.it}
\author[aff1]{Long Chen\orcidlink{0000-0002-1242-7082}}
\author[aff2,aff3]{Tommaso Dorigo\orcidlink{0000-0002-1659-8727}}
\author[aff1]{Nicolas R. Gauger\orcidlink{0000-0002-5863-7384}}
\author[aff6]{Pietro Vischia \orcidlink{0000-0002-7088-8557}}
\author[aff3,aff4,aff5]{Federico Nardi\orcidlink{0000-0003-4324-2811}}
\author[aff2,aff3]{Muhammad Awais \orcidlink{0009-0001-3665-9507}}
\author[aff7]{Hamza Hanif\orcidlink{0000-0002-0984-7887}}
\author[aff3,aff8]{Shahzaib Abbas\orcidlink{0009-0009-0346-2466}}
\author[aff9]{Rukshak Kapoor}

\affiliation[aff1]{organization={Chair for Scientific Computing, RPTU University Kaiserslautern-Landau}, country={Germany}}
\affiliation[aff2]{organization={Luleå University of Technology}, country={Sweden}}
\affiliation[aff3]{organization={INFN, sezione di Padova}, country={Italy}}
\affiliation[aff4]{organization={Università di Padova, dipartimento di Fisica e Astronomia}, country={Italy}}
\affiliation[aff5]{organization={Laboratoire de Physique Clermont Auvergne}, country={France}}
\affiliation[aff6]{organization={Universidad de Oviedo, Department of Physics and ICTEA}, country={Spain}}
\affiliation[aff7]{organization={Simon Fraser University}, country={Canada}}
\affiliation[aff8]{organization={University of Karachi}, country={Pakistan}}
\affiliation[aff9]{organization={Thapar Institute of Engineering $\&$ Technology (TIET), Patiala}, country={India}}

\begin{abstract}
In this work, we present a conditional denoising-diffusion surrogate for electromagnetic calorimeter showers that is trained to generate high-fidelity energy-deposition maps conditioned on key detector and beam parameters. The model employs efficient inference using Denoising Diffusion Implicit Model sampling and is pre-trained on \textsc{GEANT4} simulations before being adapted to a new calorimeter geometry through Low-Rank Adaptation, requiring only a small post-training dataset.

    We evaluate physically meaningful observables, including total deposited energy, energy-weighted radius, and shower dispersion, obtaining relative root mean square error values below 2$\%$  for representative high-energy cases. This is in line with state-of-the-art calorimeter surrogates which report comparable fidelity on high-level observables. Furthermore, we compare gradients of a reconstruction-based utility function with respect to design parameters between the surrogate and finite-difference references. The diffusion surrogate reproduces the qualitative structure and directional trends of the true utility landscape, providing usable sensitivities for gradient-based optimization. These results show that diffusion-based surrogates can accelerate simulation-driven detector design while enabling differentiable, gradient-informed analysis.

\end{abstract}




\end{frontmatter}



\section{Introduction}
\label{sec:intro}

Accurate simulation of particle showers in calorimeters is essential for detector design and performance optimization in high-energy physics (HEP). 
Tools such as \textsc{GEANT4} \cite{Agos2003Geant4, Allison2006, Allison2016} provide detailed and reliable physics-based simulations, but they are computationally expensive and inherently non-differentiable. 
This limits their use in modern design optimization workflows, where differentiable surrogates are required to enable gradient-based optimization and sensitivity analysis \cite{dorigo2023e2e, aehle2025differentiable}.

As the complexity of calorimeter systems increases, particularly in the context of the High-Luminosity Large Hadron Collider (HL-LHC) and future collider experiments, the number of design parameters grows substantially, encompassing variations in geometry, materials, and granularity \cite{DeVita2025HadronID}. Although gradient-free, sample-efficient optimization methods such as Bayesian optimization and evolutionary algorithms are widely used, their efficiency deteriorates rapidly with increasing dimensionality~\cite{adelmann2022surrogate}. This makes exhaustive exploration of the high-dimensional detector design space computationally challenging, motivating the development of differentiable surrogate models that can provide analytic gradients for efficient optimization. Moreover, black-box optimization approaches \cite{forrester2010black} relying solely on forward simulations offer limited interpretability and cannot provide analytic gradients of performance metrics with respect to detector design parameters.

While such methods can optimize design objectives through repeated evaluations, they remain fundamentally gradient-free, making end-to-end optimization inefficient in high-dimensional parameter spaces. In contrast, a differentiable, physics-aware surrogate model would enable end-to-end detector co-design workflows, where performance objectives such as energy resolution or shower containment can be directly optimized with respect to detector geometry and material configuration. Establishing such differentiable surrogates is non-trivial, since it requires learning a smooth and physically consistent mapping from design parameters to simulated detector responses. Achieving this differentiability forms a key step toward fully end-to-end detector co-design workflows.

To address above these challenges, we propose a diffusion-based differentiable surrogate framework for calorimeter simulation. Diffusion models (DMs) \cite{sohl2015deep, ho2020denoising} have demonstrated exceptional generative fidelity across domains and can be
conditioned on both discrete and continuous variables \cite{dhariwal2021diffusion}, making them  particularly suitable
for representing calorimeter responses under varying detector configurations. Several works have applied diffusion models to calorimeter simulation, achieving high-fidelity and fast shower generation \cite{mikuni2022score, buhmann2023caloclouds, amram2023geometry, favaro2025calodream}. Recent work have also explored pre-training and adaptation to improve generalization across detector configurations \cite{raikwar2025calodit, gaede2025crossgeometry}.

The primary objective of this work is to develop and evaluate a differentiable surrogate model for calorimeter simulations based on diffusion models. Our key contribution lies in a two-stage modeling framework that combines global generalization and local accuracy. This design is motivated by the high-dimensional nature of the detector design space, which requires a diffusion model capable of generating showers conditioned on diverse configurations. However, training such a model to cover the full design space would demand prohibitively large amounts of data and computational resources. To address this, we adopt a two-stage strategy consisting of pre-training followed by targeted post-training. In the pre-training phase, the model learns broad correlations between detector configurations and calorimeter responses across a wide range of design parameters, establishing a global representation of the simulation space.
In the post-training phase, we apply lightweight fine-tuning via Low-Rank Adaptation (LoRA) \cite{hu2021lora} to specialize the surrogate for a specific detector configuration, achieving locally accurate predictions and reliable gradients without retraining the full model. This adaptation is computationally efficient and requires only limited additional data, making it well-suited for rapid specialization to new detector configurations. This two-stage approach demonstrates how diffusion-based surrogates can provide both efficient large-scale coverage and precise local sensitivity, an essential capability for future end-to-end, gradient-based detector design pipelines. An earlier version of this work, introducing the pre- and post-training framework, was presented at the Fifth MODE Workshop on Differentiable Programming for Experiment Design \cite{MODE2025}, where the approach was developed independently with a focus on optimization-oriented surrogate modeling.

To further accelerate both training and inference, we employ a deterministic diffusion formulation \cite{song2021ddim}, which enables deterministic and differentiable sampling in significantly fewer steps compared to the Denoising Diffusion Probabilistic Model (DDPM) \cite{ho2020denoising}. 
This design facilitates end-to-end gradient propagation from the surrogate outputs to the conditioning variables, paving the way for gradient-based calorimeter optimization.

Although our framework enables the construction of a fully differentiable end-to-end optimization pipeline, linking detector design parameters to physics-level utilities through backpropagation, this paper does not aim to perform the optimization itself. Rather, we focus on establishing and validating the differentiable components that make such a pipeline feasible.

\section{Related Work}

Pre-training followed by targeted fine-tuning has been a dominant paradigm in modern machine learning, enabling models to learn broadly useful representations on large, diverse datasets and then adapt efficiently to downstream tasks with limited data \cite{girshick2014rcnn,sermanet2013overfeat,he2018rethinking, devlin2018bert}. 
This strategy has been crucial in computer vision and natural language processing \cite{houlsby2019parameter}, and recent adapter-based methods have made fine-tuning large pre-trained foundation models both parameter- and memory-efficient \cite{he2017maskrcnn, mahajan2018weakly, sun2017unreasonable}. 
Such transfer and adapter techniques motivate our two-stage pre-train/post-train strategy: a broadly trained surrogate that can be rapidly specialized to a target detector configuration with a small amount of high-fidelity data.

In HEP, a growing body of work explores machine-learned generative surrogates as fast alternatives to fast complements to first-principle based simulations like \textsc{GEANT4}. 
Early successes include GAN-based \cite{goodfellow2014generative} approaches such as \textsc{CaloGAN} \cite{paganini2018calogan}, which demonstrated orders-of-magnitude speedups for electromagnetic shower emulation while capturing several key shower features. 
Normalizing-flow approaches (e.g., \textsc{Calo\-Flow} and \textsc{Calo\-Flow~v2}) \cite{krause2023caloflow,krause2021caloflowv2}) later improved sample fidelity and offered an alternative probabilistic modeling framework for calorimeter showers. 
More recently, diffusion and score-based models have emerged as state-of-the-art generative models in other domains and are being adapted for HEP surrogate modeling \cite{adelmann2022surrogate} and optimization because of their strong sample quality and flexible conditioning capabilities \cite{ho2020denoising, song2021ddim}. 
Notably, an end-to-end detector optimization pipeline based on conditional diffusion surrogates has recently been proposed and shown in case studies for sampling calorimeters \cite{schmidt2025aido}, illustrating the practicality of differentiable generative surrogates for design exploration.

Complementary to surrogate modeling, efforts to make the underlying physics simulators themselves differentiable are also advancing. 
Algorithmic differentiation (AD) ~\cite{Aehle2022ForwardAD, Aehle2022ReverseAD, strong2024tomopt} applied to \textsc{GEANT4} has been explored to obtain pathwise derivatives of simulation outputs with respect to inputs, enabling gradient-based analyses directly on the Monte Carlo simulation chain \cite{aehle2024adgeant4, aehle2024pathwise}. 
These works demonstrate the promise of differentiable, physics-aware simulation; however, AD-enabled \textsc{GEANT4} is still computationally demanding and targeted at specific use cases, motivating surrogate-based approaches that trade a small modelling error for large gains in speed and gradient accessibility.

\section{Methodology}

Our contribution sits at the intersection of these strands described in the previous Section. 
Unlike prior GAN- or flow-based surrogates, we build a conditional, differentiable diffusion surrogate that is pre-trained over a broad design space and then efficiently specialized using LoRA-style \cite{hu2021lora} adapters; this combination leverages (i) the generative fidelity and conditioning flexibility of diffusion models, (ii) pre-training for broad generalization, and (iii) adapter-based post-training for low-cost specialization. 
In doing so we aim to provide a practical, scalable path toward surrogate models that are both high-fidelity and amenable to gradient-based design sensitivity analysis.

\subsection{Diffusion-based Surrogate Framework}

We build upon the DDPM \cite{ho2020denoising}, which iteratively learns to reverse a noise corruption process, enabling the generation of realistic samples conditioned on external variables. In our case, the model learns the conditional distribution $p_{\theta}(x|y)$, where $x$ denotes the calorimeter shower and $y$ encodes calorimeter conditioning variables (initial energy and cell size configurations).

During training, the model follows the DDPM noise-prediction objective. Given a clean calorimeter shower $x_0$, Gaussian noise $\epsilon \sim \mathcal{N}(0,I)$ is progressively added through the forward diffusion process to produce a noisy sample $x_t$ at timestep $t$:

\begin{equation}
x_t = \sqrt{\bar{\alpha}_t}x_0 + \sqrt{1-\bar{\alpha}_t}\epsilon ,
\end{equation}
where $\bar{\alpha}_t$ is determined by a predefined variance schedule.

The neural network $\epsilon_\theta(x_t,t,y)$ is trained to predict the injected noise conditioned on the detector parameters $y$ (energy, cell size, and material). The training objective minimizes the mean squared error between the predicted and true noise:

\begin{equation}
\mathcal{L} =
\mathbb{E}_{x_0,t,\epsilon}
\left[
\|\epsilon - \epsilon_\theta(x_t,t,y)\|^2
\right].
\end{equation}

By learning to estimate the noise at each diffusion step, the model captures the underlying distribution of calorimeter showers and enables sample generation through the reverse diffusion process.

To accelerate generation while maintaining fidelity, we adopt the
Denoising Diffusion Implicit Model (DDIM) sampling strategy
\cite{song2021ddim}, an alternative deterministic sampling procedure
that can be applied to models trained with the DDPM objective.
Starting from Gaussian noise $x_T \sim \mathcal{N}(0,I)$, the method
iteratively reconstructs the data sample using the learned noise
predictor while allowing generation in significantly fewer steps compared to
the original DDPM formulation. In DDIM sampling, the update from
step $t$ to $t-1$ can be written as

\begin{equation}
x_{t-1} =
\sqrt{\bar{\alpha}_{t-1}}\hat{x}_0 +
\sqrt{1-\bar{\alpha}_{t-1}}\,\epsilon_\theta(x_t,t,y),
\end{equation}
where $\hat{x}_0$ is the predicted clean sample derived from the
noise estimator $\epsilon_\theta$. This framework serves as the basis
for the conditional U-Net architecture and gradient-based optimization
described in the following sections.

\subsection{Model Architecture}

In this paper, we use a conditional diffusion model for simulating calorimeter showers, where the model learns the energy deposit process based on multiple cell size configurations. The architecture consists of a U-Net model \cite{ron2015Unet}. The network follows the standard encoder–bottleneck–decoder structure with skip connections, augmented with two conditioning mechanisms: (i) diffusion time embeddings, and (ii) calorimeter-specific embeddings encoding energy levels, cell sizes, and material properties (detailed in Sec. \ref{subsec:data}). 

Discrete conditions (binned energy) are encoded via learned embeddings, while continuous conditions (cell sizes, materials) are processed by fully connected layers. The material type is represented by a scalar label (0 for $PbF_{2}$ and 1 for $PbWO_{4}$) and provided as a conditioning input to the network. The resulting vectors are summed with the time embedding and injected into each residual block. The encoder progressively downsamples the noisy shower image, the bottleneck incorporates self-attention to capture global correlations, and the decoder upsamples while integrating encoder features to reconstruct the denoised image. 

\begin{figure}[H]
    \centering
    \includegraphics[width=\linewidth]{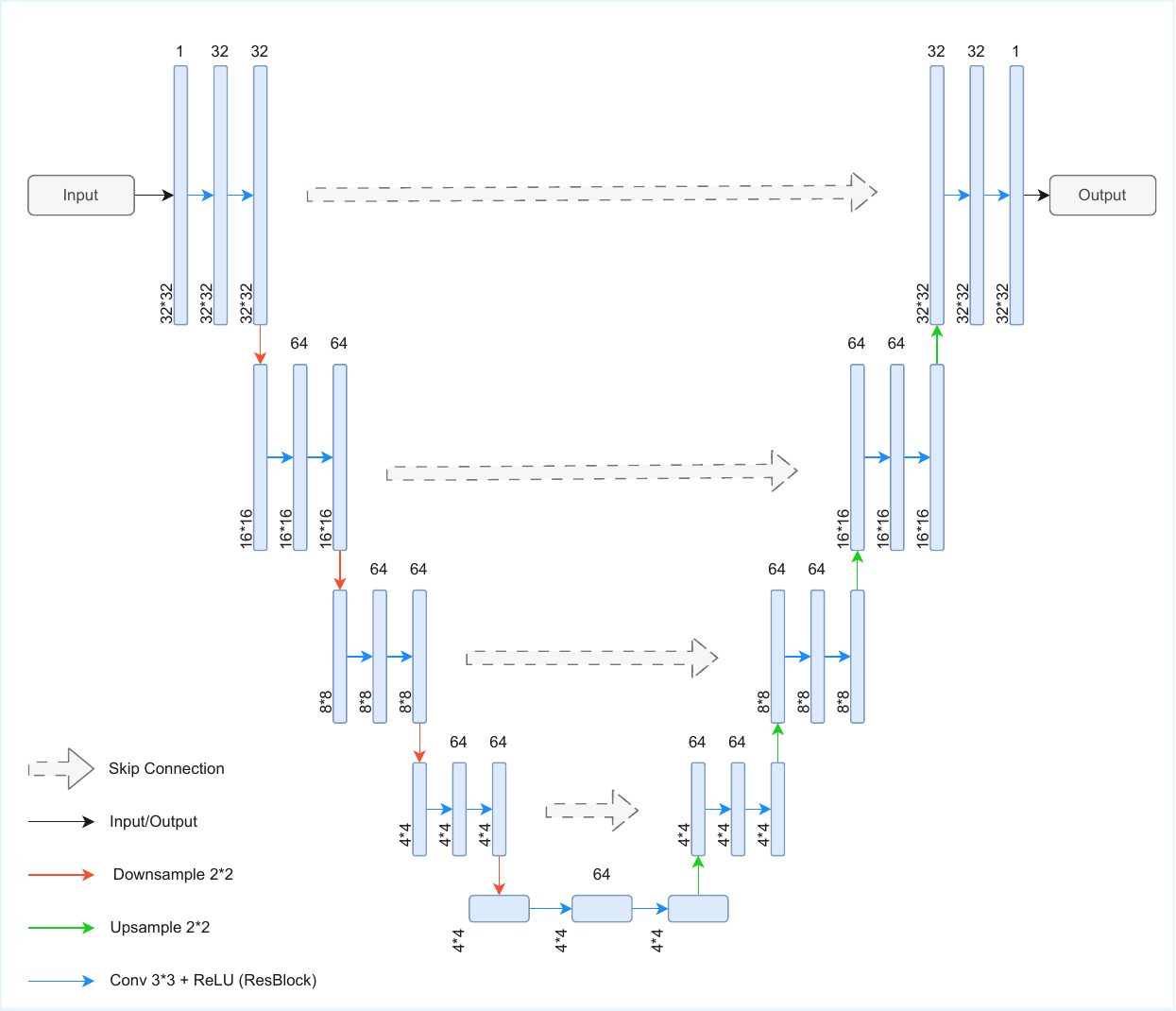}
    \caption{Architecture of the proposed conditional diffusion model for calorimeter shower simulation. The model employs a U-Net backbone that processes noisy shower images, with skip connections linking encoder and decoder stages. Calorimeter-specific conditions (energy, cell sizes, and material) together with diffusion time steps are embedded and injected into each residual block, enabling the network to learn energy deposition patterns across detector configurations. Numbers above each block indicate the number of feature maps, while numbers below indicate the spatial resolution of the feature maps.}
    \label{fig:Unet}
\end{figure}

This design enables the model to learn spatial deposition patterns under varying detector configurations while preserving fine-grained details. An overview of the architecture is shown in Fig.~\ref{fig:Unet}.

\subsection{Low-Rank Adaptation (LoRA)}
To make the model's capacity more flexible and efficient for a specific detector configuration
, we incorporate LoRA adapters \cite{hu2021lora} into the convolutional layers of the U-Net architecture. The LoRA technique reduces the number of parameters required for fine-tuning by introducing low-rank projections to the existing layers. This method is particularly useful for adapting the model without retraining all the parameters from scratch.  

\begin{figure}[H] 
    \centering
    \includegraphics[width=\linewidth]{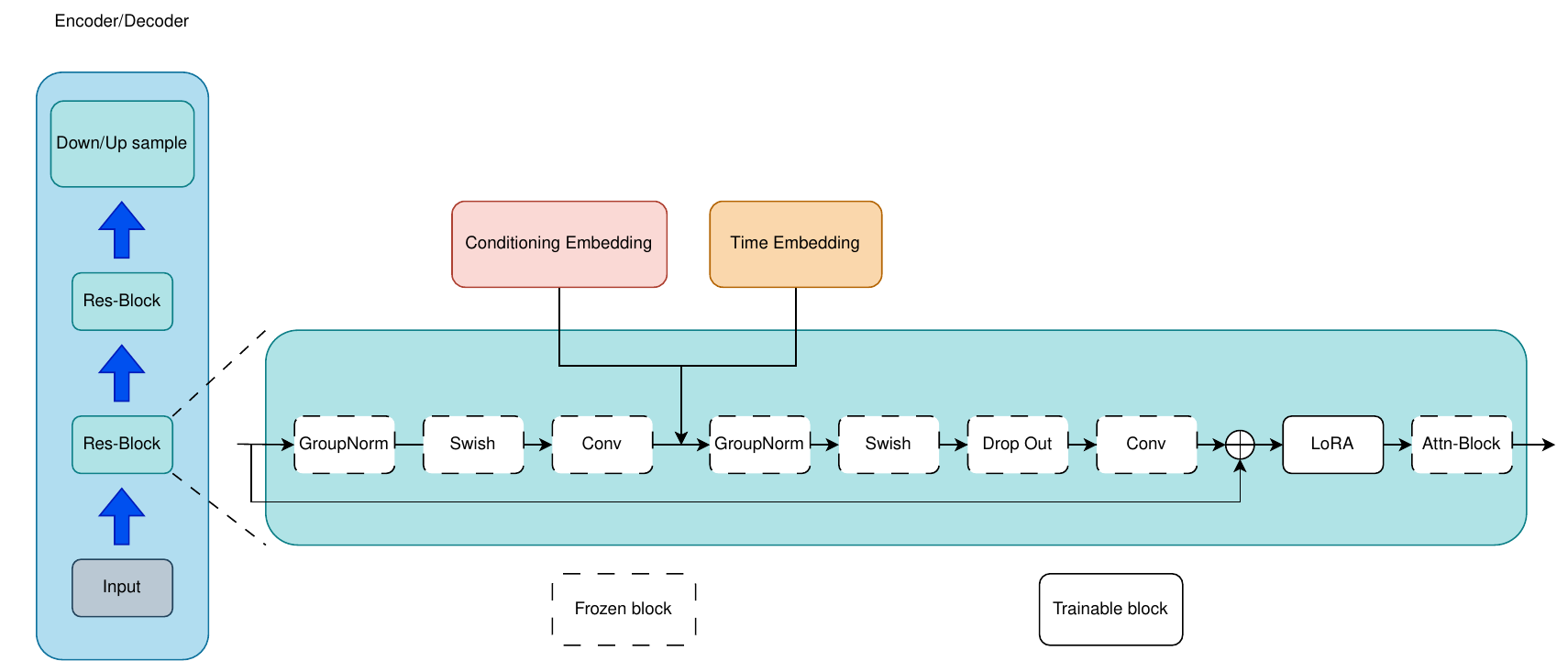}
    \caption{Schematic of a ResBlock in the conditional U-Net architecture with LoRA adapters. The input feature map passes through the first convolution block (GroupNorm → Swish → Conv), then time and conditioning embeddings are added. The second convolution block (GroupNorm → Swish → Dropout → Conv) follows, after which the residual shortcut and LoRA adapter are applied.}
    \label{fig:resblock_lora_scheme}
\end{figure}

In our implementation, LoRA adapters are attached to the convolutional layers within each Residual Block (ResBlock) in the downsampling, middle, and upsampling paths. During fine-tuning, the original weights are frozen, ensuring that only the parameters of the low-rank matrices are updated. This selective training significantly reduces the number of trainable parameters and allows for the efficient adaptation of the model to the specific use case without compromising the stability of the pre-trained network (see Fig.~\ref{fig:resblock_lora_scheme}).

\subsection{Gradient Formulation and Sensitivity Analysis}
\label{subsec:grad_analysis}

The conditional diffusion surrogate defines a differentiable mapping  $f_{\theta}: y \rightarrow x$, where $y$ represents detector design parameters (such as cell sizes, materials, and incoming energy), and $x$ denotes the corresponding calorimeter shower. Once trained, the model approximates the conditional distribution $p_{\theta}(x|y)$, enabling the generation of physically consistent showers for arbitrary design configurations.

The differentiability of $f_{\theta}$ allows gradients to propagate through the surrogate to perform gradient-based detector optimization. A differentiable objective function $L(y)$ representing a design figure of merit, such as energy resolution, response linearity, or shower containment, can be defined over generated samples. The gradients $\nabla_{y} L(y)$ can then be computed via automatic differentiation and used to update the detector parameters iteratively:
\begin{equation}
 y_{k+1} = y_k - \lambda \nabla_y L(y_k)
\end{equation}
where $\lambda$ is the learning rate that controls the optimization step size.

\section{Experiments}

\subsection{Considered use case scenario}
\label{subsec:scenario}
We apply the model described in the previous section to the reconstruction of high-energy photons in an electromagnetic calorimeter for a future muon collider detector\cite{Bartosik2022, accettura2023, Bartosik2020MuonCollider}. In a muon collider, muon decays in the beam produce a large flux of particles (mostly soft photons and neutrons, see Fig.~\ref{fig:calo_config}) that impact the detector facility from the sides. A significant beam-induced background (BIB) \cite{mokhov2012detector, collamati2021} thus impinges in the inner layers of the electromagnetic calorimeter. Preserving the performance of the electromagnetic calorimeter in the presence of the BIB is a non-trivial task, which calls for both purely hardware solutions, such a tungsten nozzle that is the subject of an independent optimization study \cite{Cera2022Crilin} and studies like the one described here. In this application, the per-cell background term $E_{bgr}$ introduced in Sec.~\ref{subsec:grad_analysis}  corresponds to the energy deposited by this BIB. The geometry of the calorimeter cells has, in the considered application, a significant potential to affect the energy and position resolution of energetic gamma showers; in this work we are concerned with a precise differentiable modeling of the energy deposition from the EM showers in the presence of BIB. 

The gradient and sensitivity experiments reported in Sec.~\ref{subsec:gradient} we employ a
calorimeter environment representative of the muon-collider design scenario. Muon decays in the collider produce a diffuse, depth-biased and highly stochastic BIB which can strongly affect calorimeter observables. To evaluate the surrogate under realistic conditions we overlay per-cell BIB maps onto signal showers prior to reconstruction and utility computation. Where available, BIB maps derived from machine-detector-interface (MDI) simulations (e.g., FLUKA/MARS \cite{ferrari2005fluka, mokhov2017mars15} particle fluences propagated through the detector via \textsc{GEANT4}) were resampled to the voxelization used in this study. When MDI-derived maps were not available for a given geometry, we used a validated parametric emulator calibrated to reproduce the principal BIB features (depth-dependent profile, spatial diffusion across transverse cells, and large event-to-event variance). For each event the overlay procedure samples a per-cell background map (either MDI-derived or emulator-generated) and adds it to the per-cell signal energies before applying the reconstruction mask (Eq.~\ref{eq:wi_main}–\ref{eq:Emeas_main}) and computing the utility.

To demonstrate this framework, we compute the gradients of a downstream utility with respect to the per-cell signal outputs produced by the generative surrogate (see Sec.~\ref{subsec:scenario} for the test setup). For each event and each
cell \(i\) we denote the model outputs for signal and background by \(E_{\mathrm{sig},i}\) and \(E_{\mathrm{bgr},i}\), respectively. 
The per-cell total energy is then
\begin{equation}
E_i \;=\; E_{\mathrm{sig},i} + E_{\mathrm{bgr},i}.
\end{equation}

We apply a differentiable soft mask to discriminate signal-like cells from the background:
\begin{equation}\label{eq:wi_main}
w_i \;=\; \sigma\big(K(E_i-T)\big)
\;=\; \frac{1}{1+\exp\big(-K(E_i-T)\big)},
\end{equation}
where \(T\) is a threshold (we use a background-calibration mean plus one standard deviation unless otherwise noted) and \(K\) scales the sharpness of the mask. The reconstructed energy used in the utility is the masked sum.
\begin{equation}\label{eq:Emeas_main}
E_{\mathrm{meas}} \;=\; \sum_{i=1}^{N_{\mathrm{cell}}} w_i\, E_i.
\end{equation}

We define the utility to be maximized as the inverse of a stabilized MSE:
\begin{equation}\label{eq:U_main}
U \;=\; \frac{1}{\alpha^2 + \frac{1}{N_{\mathrm{ev}}}\sum_{n=1}^{N_{\mathrm{ev}}}\big(E_{\mathrm{meas}}^{(n)}-E_{\mathrm{true}}^{(n)}\big)^2 }.
\end{equation}
where $\mathrm{MSE}$ denotes the per-event mean-squared error between the DDPM-generated shower and the corresponding \textsc{GEANT4} reference, and $\alpha>0$ is a small regularization constant. This choice was motivated by two practical considerations. This inverse MSE form was chosen as a simple, monotonic proxy for sample fidelity: improvements in agreement increase $U$, while degradations decrease it, making the gradient direction interpretable for design optimization. Although not a physics-driven objective and potentially sensitive when the MSE becomes very small, it provides a stable and convenient test function for assessing differentiability and comparing DDPM-derived gradients with finite-difference (FD) references. More physically motivated utility definitions are left for future work.

\begin{figure}[H] 
    \centering
    \includegraphics[width=\linewidth]{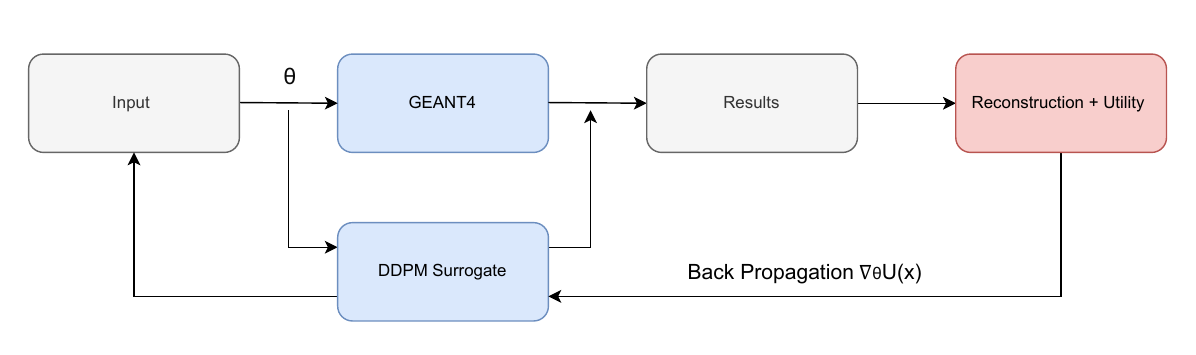}
    \caption{Schematic of the differentiable reconstruction–utility pipeline. The forward path flows from the design parameters $\theta$ into either the \textsc{GEANT4} simulation or the DDPM surrogate, producing calorimeter results that are passed to the reconstruction and utility function. The backward path illustrates gradient backpropagation of the utility $U(x)$ with respect to the design parameters $\theta$, which occurs only through the differentiable DDPM surrogate.}
    \label{fig:grad_flow_scheme}
\end{figure}

Equation~\ref{eq:wi_main}–\ref{eq:U_main} defines a differentiable reconstruction–utility chain that connects the simulated calorimeter response to a scalar design objective. Since the surrogate diffusion model $f_{\theta}(y)$ is differentiable with respect to its conditioning variables, the gradient of the utility $U$ with respect to the design parameters can be obtained by backpropagating through the entire sampling and reconstruction process. 
As shown in Fig.~\ref{fig:grad_flow_scheme}, the end-to-end differentiable pipeline consists of the \textsc{GEANT4}-based simulation and conditioning parameters $\theta$, the diffusion surrogate, reconstruction, and the resulting gradient flow that propagates back to the design variables. 

This gradient $\nabla_\theta U(x)$ quantifies the sensitivity of the chosen physics performance metric to each detector design parameter, thereby enabling gradient-based optimization of calorimeter geometry, segmentation, and material configuration.

\subsection{Dataset and Setup}
\label{subsec:data}

\begin{figure}[H] 
    \centering
    \includegraphics[width=\linewidth]{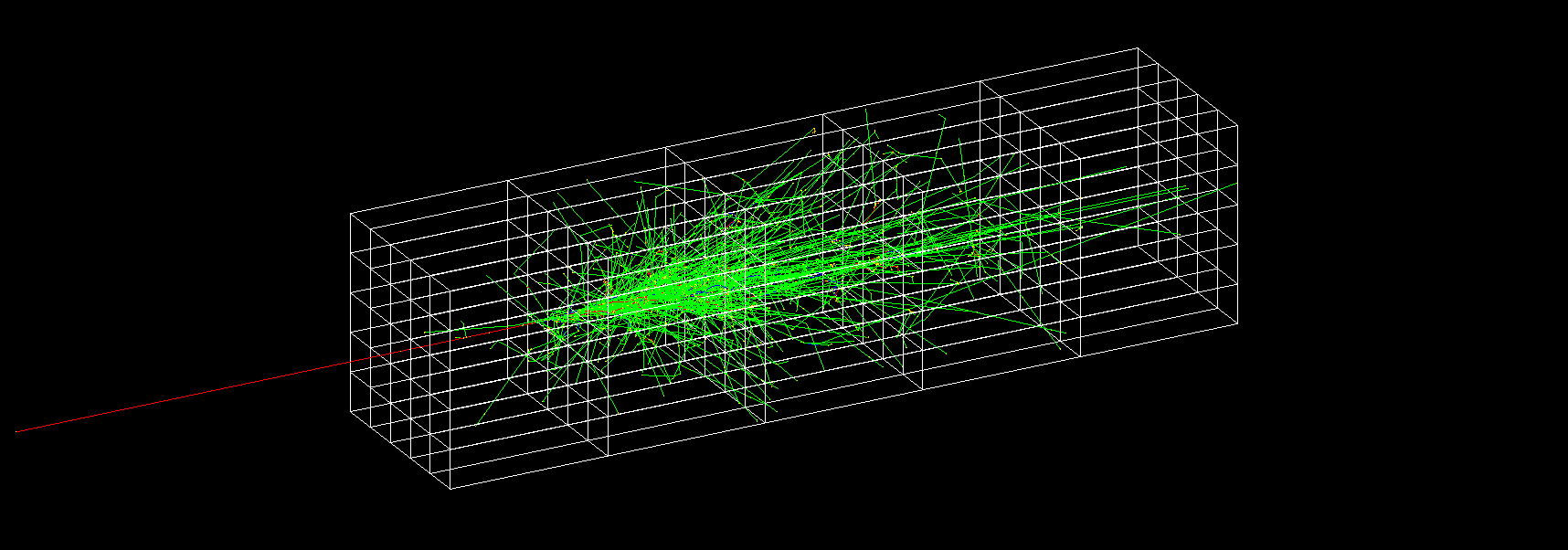}
    \caption{\textsc{GEANT4} visualization of a 4 $\times$ 4 $\times$ 10 $\mathrm{cm}^3$ $PbF_2$ scintillator cell used in the pre-training dataset. The incident photon (red line) initiates an electromagnetic shower whose energy deposits are recorded in the voxelized calorimeter grid (white). Green tracks represent secondary particles generated during the cascade.}
    \label{fig:calo_config}
\end{figure}

The pre-training dataset was produced using \textsc{GEANT4} (version 11.3.0) with the \textsc{FTFP\_BERT} physics list on a nominal $5\times5\times5$ cell grid (Fig.~\ref{fig:calo_config}). We simulated a single scintillator material Lead Fluoride ($PbF_2$) across a set of five different cell-size families: $1\times1\times5$, $2\times2\times4$, $3\times3\times8$, $4\times4\times10$, and $5\times5\times15\ \mathrm{cm}^3$. For each cell configuration, eleven values of the incident photon energy were simulated (1, 10, 20, 30, 40, 50, 60, 70, 80, 90, 100\,GeV), contributing equally to a cumulative dataset of 100,000 events. Although only one material was simulated, its value was encoded as a continuous conditioning variable, allowing the model to learn material-dependent features; this pre-training dataset serves as a proof-of-concept before extending to multiple materials. The verification of good modeling properties on this set of geometries is a prerequisite for the integration of the model in a differentiable optimization pipeline for the muon collider detector.

To reduce the input dimensionality and computational cost, the 3D energy
deposition $E(x,y,z)$ from \textsc{GEANT4} is represented using two
two-dimensional energy maps corresponding to the $(x,z)$ and $(y,z)$ planes,
obtained by projecting the shower onto these planes. Energy-weighted
histograms are constructed and down-sampled to $32\times32$ pixels to form
the DDPM inputs. In this representation, the longitudinal shower development
is preserved along the $z$ axis of the input maps.  For post-training adaptation, we used 10,000 \textsc{GEANT4} events corresponding to an unseen geometry ($2.5\times2.5\times6\ \mathrm{cm}^3$) to assess LoRA fine-tuning.

\subsection{Evaluation Metrics}

To evaluate the performance of the model, we use several physical metrics that quantify the quality of the simulated calorimeter showers. These metrics provide insights into the total energy, spatial distribution, and dispersion of the energy depositions. Full definitions and implementation details are given in~\ref{app:metrics_impl}; here we report the concise forms used in the Results.

\begin{itemize}
    \item Total Energy:
    \begin{equation}\label{total_e}
        E = \sum_{x,y} I(x,y)
    \end{equation}

    \item Energy-Weighted Radius:
    \begin{equation}\label{R_E}
        R_{\text{E}} = \sqrt{\frac{\sum_{x,y} I(x,y) \left[ (x-\bar{x})^2 + (y-\bar{y})^2 \right]}{\sum_{x,y} I(x,y)}}
    \end{equation}

    \item Shower Dispersion:
    \begin{equation}\label{sigma_y}
        \sigma_{y} = \frac{\sum_{y}(y-\bar{y})^2I_{y}y}{\sum_{y}I_{y}(y)}
    \end{equation}
\end{itemize}

To quantify the error between the simulated and the ground truth for these metrics, we computed the Relative Root Mean Square Error (RRMSE) for each of the metrics.

\section{Results}
\subsection{Pre-trained model}

\begin{figure}[H] 
    \hspace{-1cm}
    \includegraphics[width=1.2\linewidth]{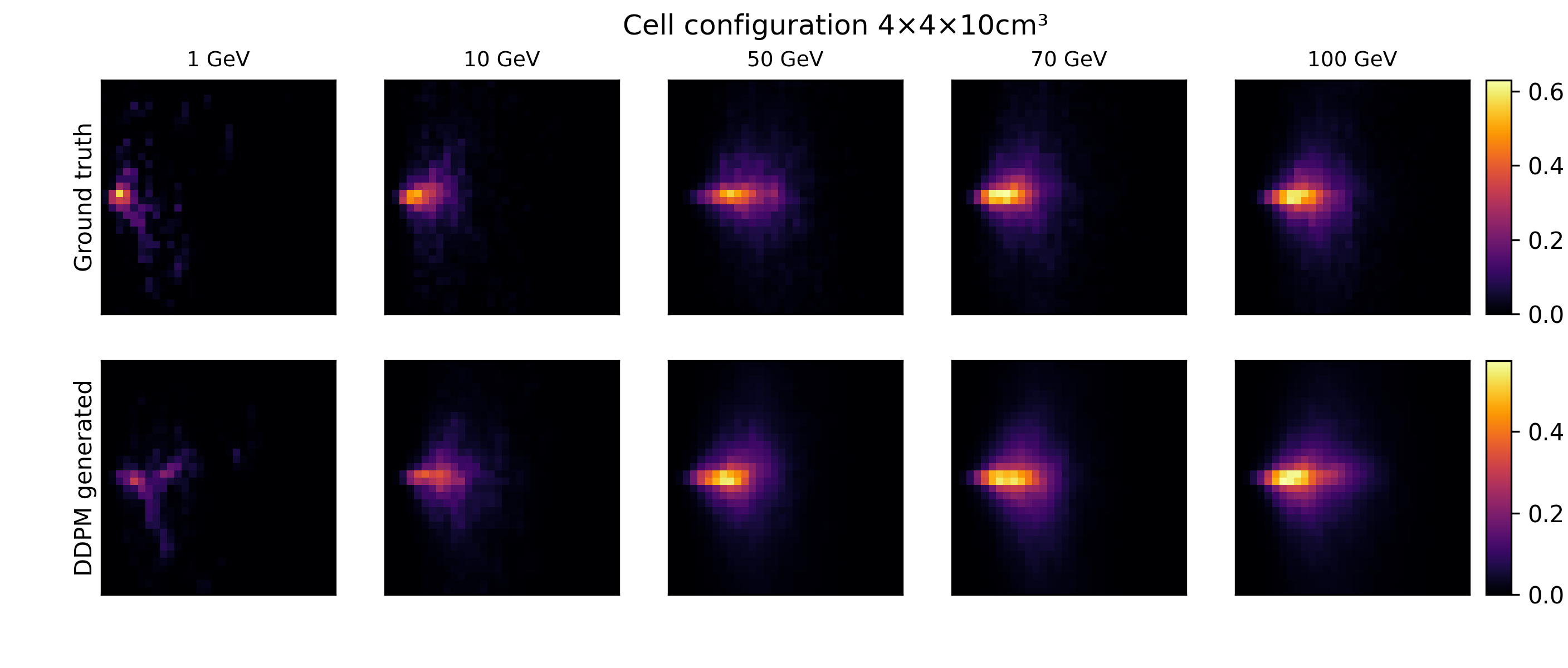}
    \caption{Comparison of Monte Carlo and diffusion-generated energy-deposition maps for a $4 \times 4 \times 10~cm^3$ $PbF_2$ scintillator cell.
    The top row shows ground-truth energy-deposition maps from the dataset, while the bottom row shows samples generated by the conditional DDPM under the same detector and beam conditions.
    Each column corresponds to a different primary photon energy.
    All images are downsampled to $32 \times 32$ pixels and displayed using the \texttt{inferno} colormap.}
    \label{fig:compare_shower}
\end{figure}

As shown in Fig.~\ref{fig:compare_shower}, the generated samples closely resemble the ground truth across all energy levels. The top row presents the ground-truth longitudinal shower profiles from \textsc{GEANT4} simulations, while the bottom row shows the profiles generated by the DDPM model for comparison. Each column corresponds to a selected primary energy of 1, 10, 50, 70, and 100 GeV, illustrating how the shower structure evolves with energy. These results indicate that the generated samples closely reproduce the ground-truth showers across all energy levels, with only subtle visual differences, confirming the DDPM’s capability to accurately model both the spatial and longitudinal development of electromagnetic showers.

\begin{figure}[H] 
    \hspace{-1cm}
    \includegraphics[width=1.2\linewidth]{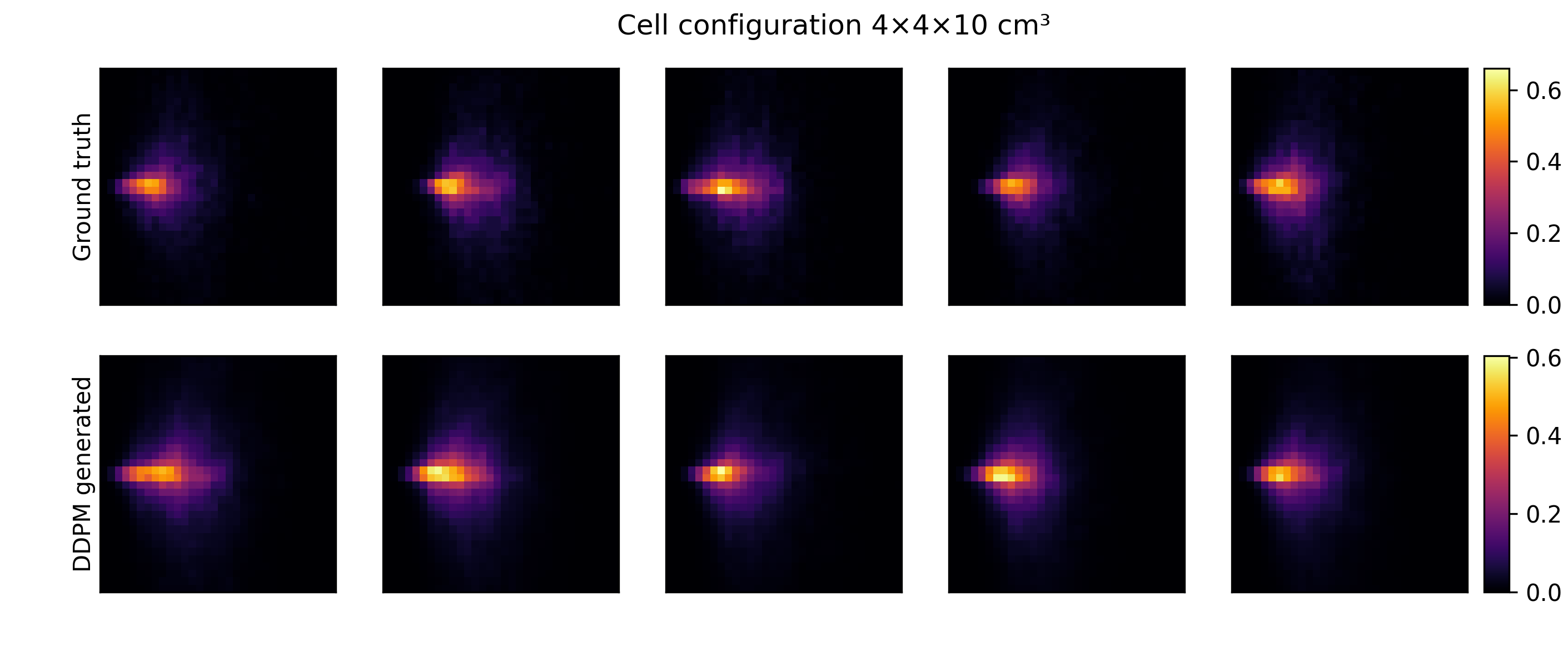}
    \caption{Comparison of Monte Carlo and diffusion-generated calorimeter showers under a single detector configuration. 
    The top row shows ground-truth energy-deposition maps from the dataset, while the bottom row shows samples generated by the conditional DDPM. 
    All panels correspond to a fixed condition of 50~GeV incident photon energy and a $4 \times 4 \times 10~cm^3$ $PbF_2$ scintillator cell. 
    Multiple samples are shown to illustrate that the diffusion model produces stochastic showers with spatial distributions similar to those of the Monte Carlo simulation. 
    All images are downsampled to $32 \times 32$ pixels and displayed using the \texttt{inferno} colormap.}
    \label{fig:compare_shower_single}
    \label{fig:compare_shower_single_condition}
\end{figure}

We further assess whether the model captures the variability of the underlying distribution. As shown in Fig.~\ref{fig:compare_shower_single_condition}, the generated samples exhibit a spread in spatial structure and intensity comparable to Monte Carlo showers, indicating that the model captures the stochastic nature of electromagnetic shower development.

\begin{figure}[H] 
    \centering
    \includegraphics[width=\linewidth]{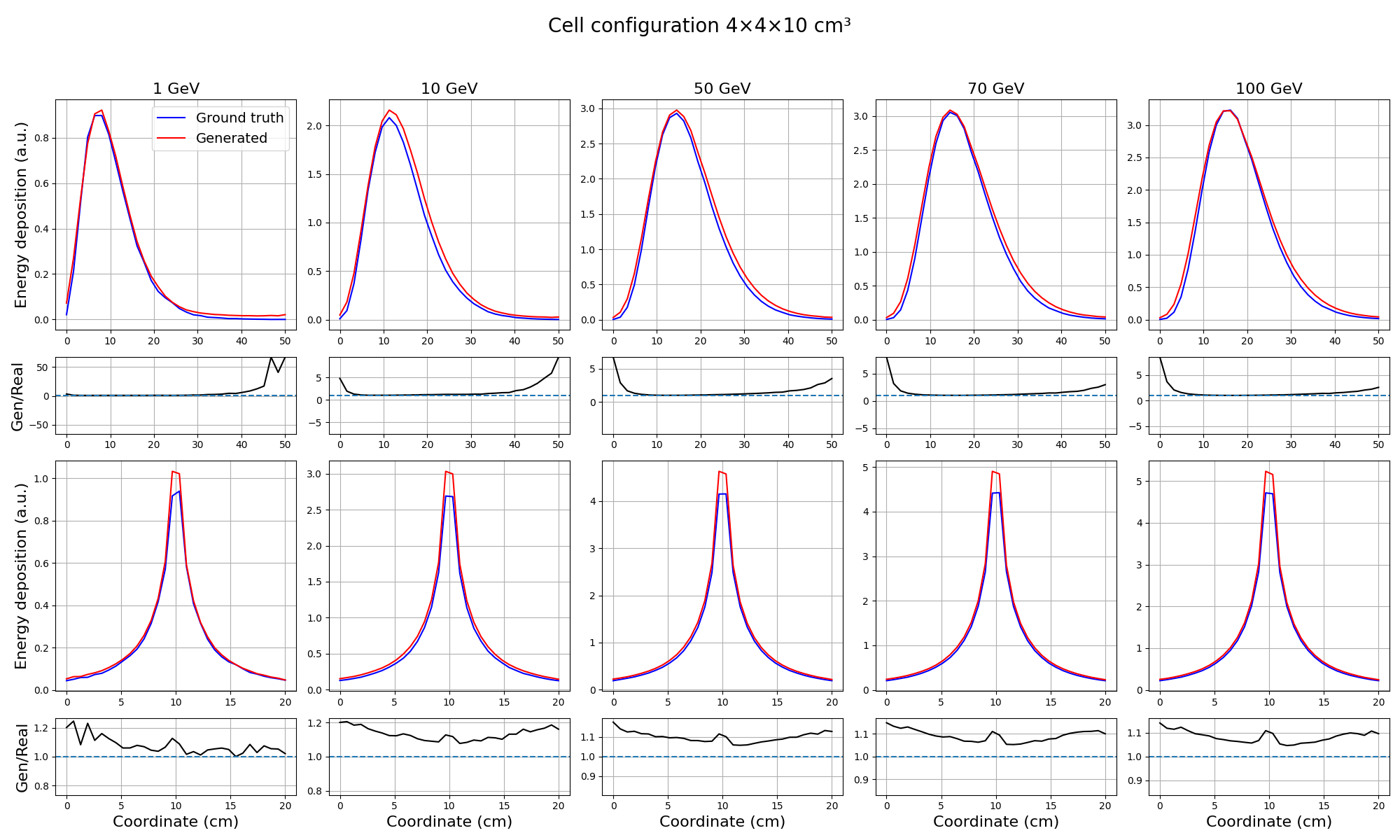}
    \caption{Longitudinal (top row) and transverse (bottom row) average energy deposition profiles for a 4 $\times$ 4 $\times$ 10 $cm^3$ $PbF_2$ scintillator cell across primary energies from 1 to 100 GeV. Blue curves denote real samples from the down-sampled \textsc{GEANT4} dataset ($32 \times 32$), while red curves indicate generated samples from our conditional DDPM model under the same (normalized) conditions. The longitudinal profile is averaged over transverse coordinates ($x$ and $y$), and the transverse profile is taken at the shower maximum in depth. Each subplot corresponds to a specific primary energy.}
    \label{fig:compare_energy_profile}
\end{figure}

To provide a more quantitative comparison, Fig.~\ref{fig:compare_energy_profile} presents the average longitudinal and transverse energy deposition profiles for a $PbF_2$ calorimeter cell with a transverse size of 4 $\times$ 4 $\times$ 10 $cm^3$. Each column corresponds to a selected primary energy of 1, 10, 50, 70, and 100 GeV. The top row displays the longitudinal profile averaged over the transverse plane, while the bottom row shows the transverse profile extracted at the shower maximum in depth. Across all energies, the generated profiles closely follow the ground truth, with slightly broader distributions at higher energies, particularly noticeable in the longitudinal component. This demonstrates the model’s capability to accurately capture both the energy dependence and spatial structure of electromagnetic showers.

\begin{figure}[H] 
    \centering
    \includegraphics[width=\linewidth]{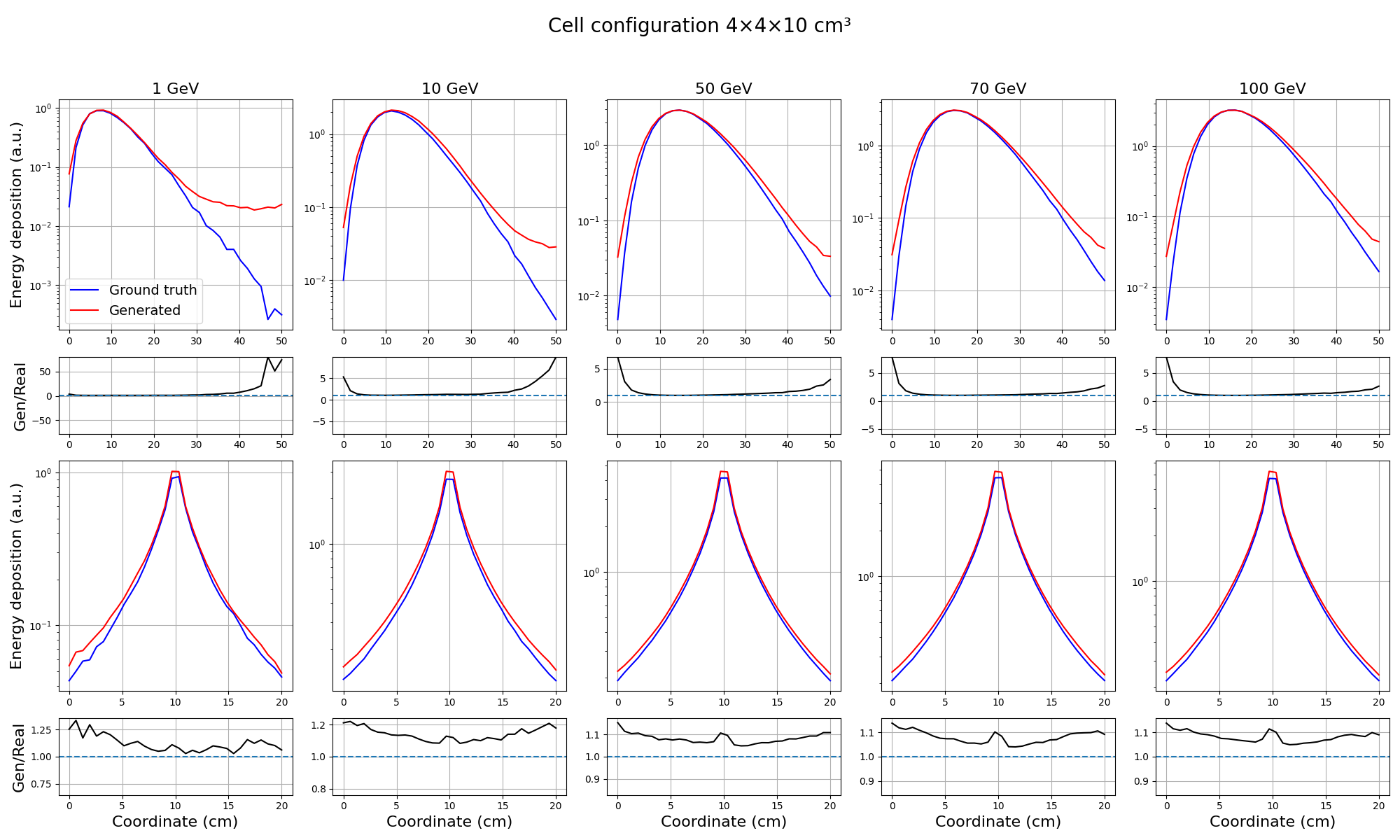}
    \caption{Longitudinal (top row) and transverse (bottom row) energy deposition profiles on a logarithmic scale for the same configuration as Fig.~\ref{fig:compare_energy_profile}. The log scale enhances the visibility of low-energy regions, enabling a more detailed comparison between ground-truth (blue) and DDPM-generated (red) distributions across primary energies.}
    \label{fig:log_compare_energy_profile}
\end{figure}

Since the model is conditioned on normalized geometry and energy parameters, this consistency across profiles reflects its ability to learn smooth, disentangled representations of these conditions. The progressive broadening of the longitudinal profile with increasing energy is particularly well captured, which aligns with physical expectations. This behavior remains consistent on a logarithmic scale (Fig.~\ref{fig:log_compare_energy_profile}), where low-energy regions are emphasized. The generated distributions closely follow the ground truth across the full dynamic range, with no significant deviations observed in the low-energy regime.

\subsection{Fidelity of Model}

To assess the fidelity of the generated calorimeter showers beyond visual similarity, we evaluate the model using physically meaningful metrics described in Eq.~\ref{total_e}–\ref{sigma_y}. These include the total deposited energy, energy-weighted shower radius $R_E$, and shower dispersion $\sigma_y$, each computed per event. The accuracy of each metric is quantified using the RRMSE defined in Eq.~\ref{app:eq:RRMSE}.

\begin{figure}[H] 
    \centering
    \includegraphics[width=\linewidth]{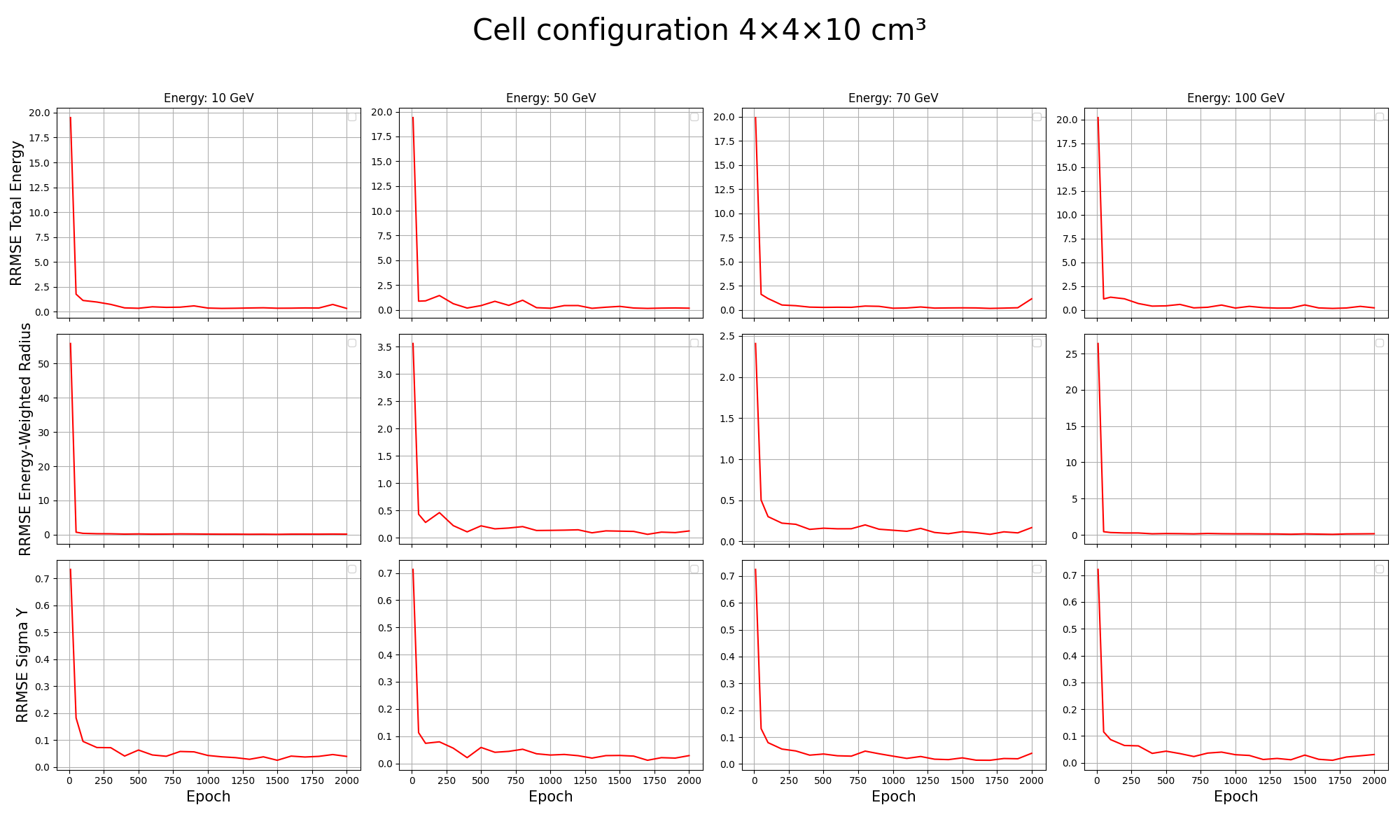}
    \caption{RRMSE as a function of training epoch for the total energy (top row), energy-weighted radius (middle row), and shower dispersion $\sigma_y$ (bottom row). Each column corresponds to a different primary energy (10, 50, 70, 100 GeV). Results are shown for the $PbF_2$ calorimeter with 4 $\times$ 4 $\times$ 10 $cm^3$. The error drops significantly during early training and stabilizes at low values, indicating good agreement with ground-truth showers across all metrics.}
    \label{fig:metrics_eval}
\end{figure}

The RRMSE is evaluated as a function of training epoch for five representative energy levels (1, 10, 50, 70, and 100 GeV), using the $PbF_2$ calorimeter configuration with transverse size 4 $\times$ 4 $\times$ 10 $cm^3$ (Fig.~\ref{fig:metrics_eval}). Each row corresponds to a different metric: total energy (top), energy-weighted radius (middle), and shower dispersion in the $y$-direction (bottom).

Across all energies and metrics, the RRMSE decreases rapidly during the early training phase and stabilizes after approximately 500 epochs. This trend confirms that the model learns to match both the global properties (total energy) and spatial characteristics (radius and spread) of the ground-truth showers. Particularly, for higher energies (e.g., 70 and 100 GeV), the model achieves an RRMSE below 2$\%$ for all three metrics, indicating robust performance even in high-energy regimes where shower fluctuations are more pronounced.

The lowest RRMSE is consistently observed for the shower dispersion $\sigma_y$, suggesting that the model reproduces the shape and width of the shower core with high accuracy. While the total energy RRMSE starts high (as expected due to random initialization), it converges quickly to below 2$\%$ across all energy levels. These results validate the model’s ability to generate physically consistent showers that match key statistical and spatial properties of real calorimeter data.

\subsection{LoRA post-training Adaptation}

To assess the generalization capabilities of our conditional DDPM model, we test its performance on a calorimeter geometry unseen during pre-training: a $PbF_2$ cell with transverse dimensions 2.5 $\times$ 2.5 $\times$ 6 $cm^3$. We compare the behavior of the original pre-trained model and the same model after post-training using LoRA. This setup allows us to isolate the impact of LoRA on model adaptation to new geometries.

\begin{figure}[H] 
    \centering
    \includegraphics[width=\linewidth]{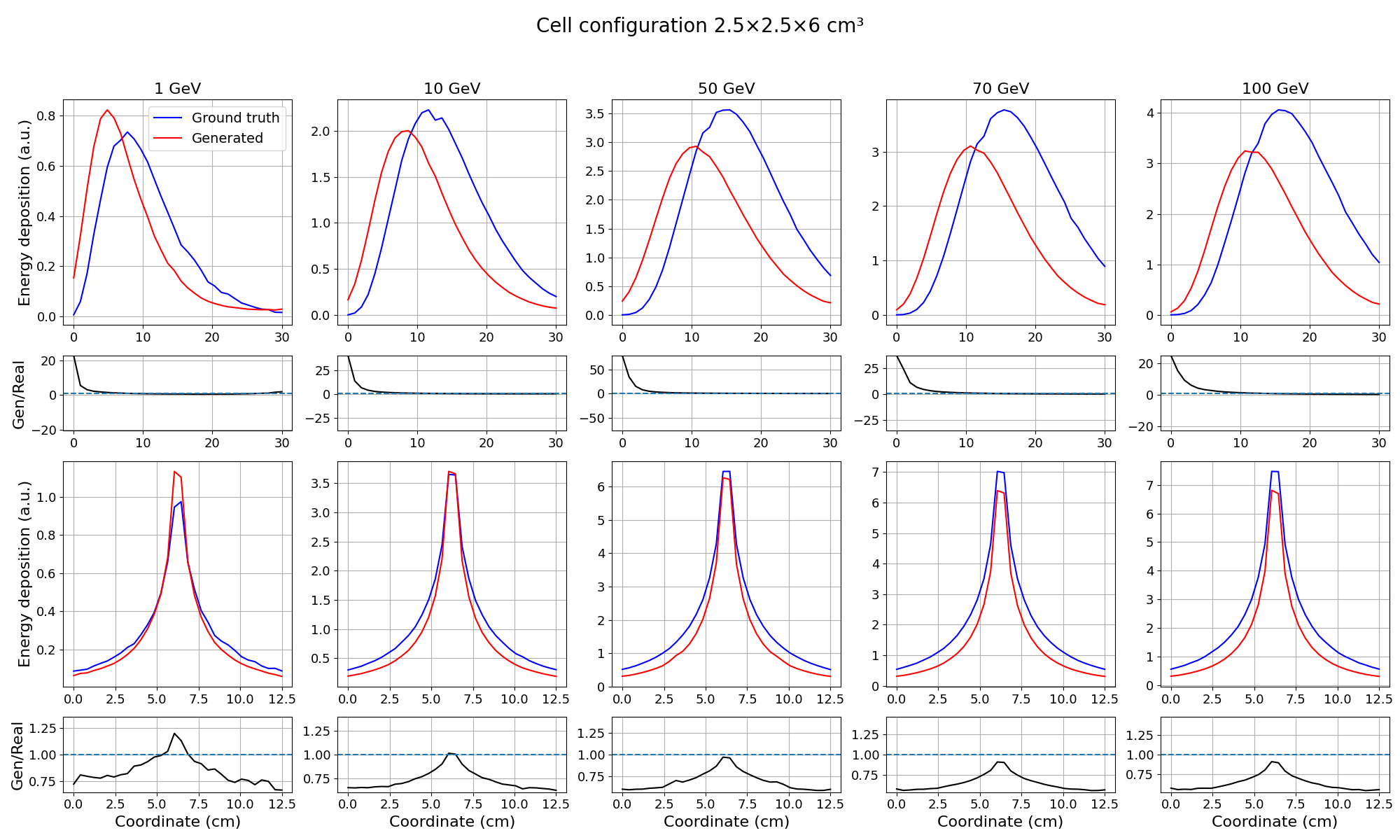}
    \caption{Longitudinal (top row) and transverse (bottom row) average energy deposition profiles for a $2.5 \times 2.5 \times 6\,\mathrm{cm}^3$ $PbF_2$ calorimeter cell, a configuration unseen during pre-training. Blue curves (ground truth) and red curves (generated) show the corresponding samples generated by the post-trained model with LoRA adaptation. Although this geometry was not part of the original training data, the model captures essential shower characteristics, especially in the transverse profiles.}
    \label{fig:unseen_pretrain}
\end{figure}

The longitudinal and transverse energy deposition profiles generated by the pre-trained model, which was not exposed to this geometry during training, are shown in Fig.~\ref{fig:unseen_pretrain}. While the transverse profiles remain reasonably close to the ground truth, the longitudinal profiles exhibit systematic underestimation across all energy levels, indicating limited generalization in depth-wise energy deposition under unseen geometry conditions.

\begin{figure}[H] 
    \centering
    \includegraphics[width=\linewidth]{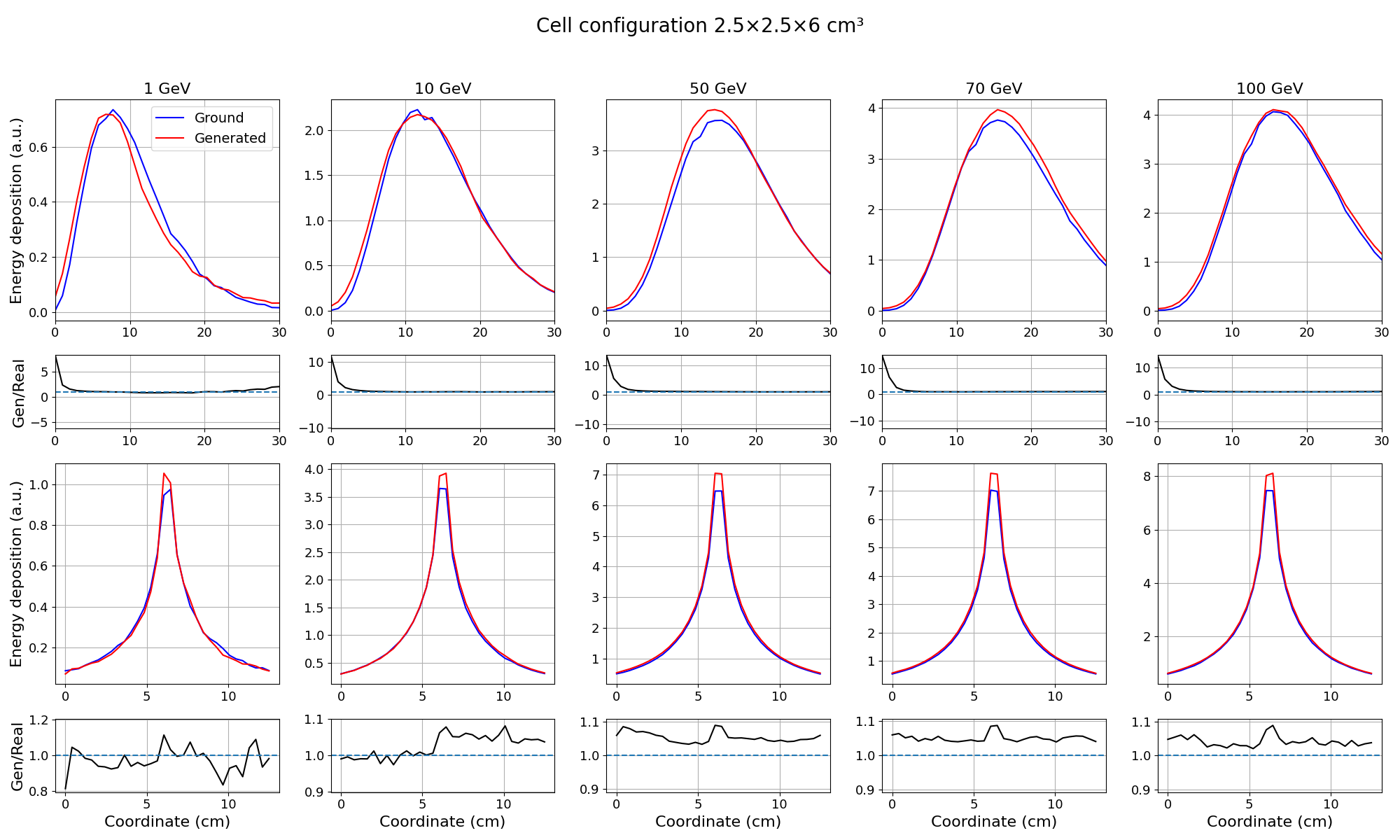}
    \caption{Longitudinal (top row) and transverse (bottom row) energy deposition profiles for the same $2.5 \times 2.5 \times 6\,\mathrm{cm}^3$ $PbF_2$ cell as in Figure~\ref{fig:unseen_pretrain}, but using the model after post-training with LoRA. The red curves (generated) now closely match the blue curves (ground truth) across all energy levels, especially in the longitudinal direction.}
    \label{fig:unseen_lora}
\end{figure}

On the other hand, Fig.~\ref{fig:unseen_lora} illustrates the same comparison for the post-trained model using LoRA, showing significant improvements. The longitudinal profiles now closely match the ground truth across the energy range, particularly in the peak region and overall shape. Transverse profiles remain accurate and even slightly improved. 

We evaluate the physical fidelity of the model after post-training using the shower observables defined in Eqs.~\ref{total_e}–\ref{sigma_y}. The metrics are computed for a detector configuration of $2.5 \times 2.5 \times 6~\mathrm{cm}^3$, not included in the original training data, and the corresponding RRMSE values for the pre-trained and post-trained models are summarized in Tab.~\ref{tab:rrmse_metrics}.

\begin{table}[htbp]
\centering
\begin{minipage}{\linewidth}
\caption{RRMSE of shower observables before and after post-processing.}
\label{tab:rrmse_metrics}

\resizebox{\linewidth}{!}{
\begin{tabular}{c cc cc cc}
\toprule
& \multicolumn{2}{c}{Total energy} & \multicolumn{2}{c}{Energy-weighted radius $R_E$} & \multicolumn{2}{c}{$\sigma_y$} \\
\cmidrule(lr){2-3} \cmidrule(lr){4-5} \cmidrule(lr){6-7}
Energy (GeV) & pre & post & pre & post & pre & post \\
\midrule
10  & 0.8379 & 0.7906 & 0.8985 & 0.8791 & 0.7127 & 0.6820 \\
20  & 0.7852 & 0.7223 & 0.8922 & 0.8769 & 0.7126 & 0.6826 \\
30  & 0.7765 & 0.6843 & 0.8978 & 0.8771 & 0.7138 & 0.6833 \\
40  & 0.7453 & 0.6527 & 0.8909 & 0.8765 & 0.7140 & 0.6840 \\
50  & 0.7265 & 0.6309 & 0.8872 & 0.8763 & 0.7138 & 0.6837 \\
60  & 0.7352 & 0.6176 & 0.8944 & 0.8770 & 0.7141 & 0.6847 \\
70  & 0.7251 & 0.5969 & 0.8927 & 0.8765 & 0.7141 & 0.6842 \\
80  & 0.7232 & 0.5905 & 0.8951 & 0.8764 & 0.7147 & 0.6845 \\
90  & 0.7102 & 0.5800 & 0.8932 & 0.8771 & 0.7146 & 0.6851 \\
100 & 0.7112 & 0.5715 & 0.8939 & 0.8768 & 0.7147 & 0.6850 \\
\bottomrule
\end{tabular}
}
\end{minipage}
\end{table}

Across all energies, the post-trained model consistently achieves lower RRMSE values, indicating improved agreement with the ground truth. The largest improvement is observed for the total energy, while the spatial metrics ($\sigma_y$ and $R_E$) show smaller but consistent gains, suggesting that the pre-trained model already captures the overall shower shape well. These results demonstrate that LoRA enables accurate adaptation of the generative model to novel detector configurations with minimal additional training.

\subsection{Gradient Computation and Analysis}
\label{subsec:gradient}

\begin{figure}[H]
\centering
\includegraphics[width=\textwidth]{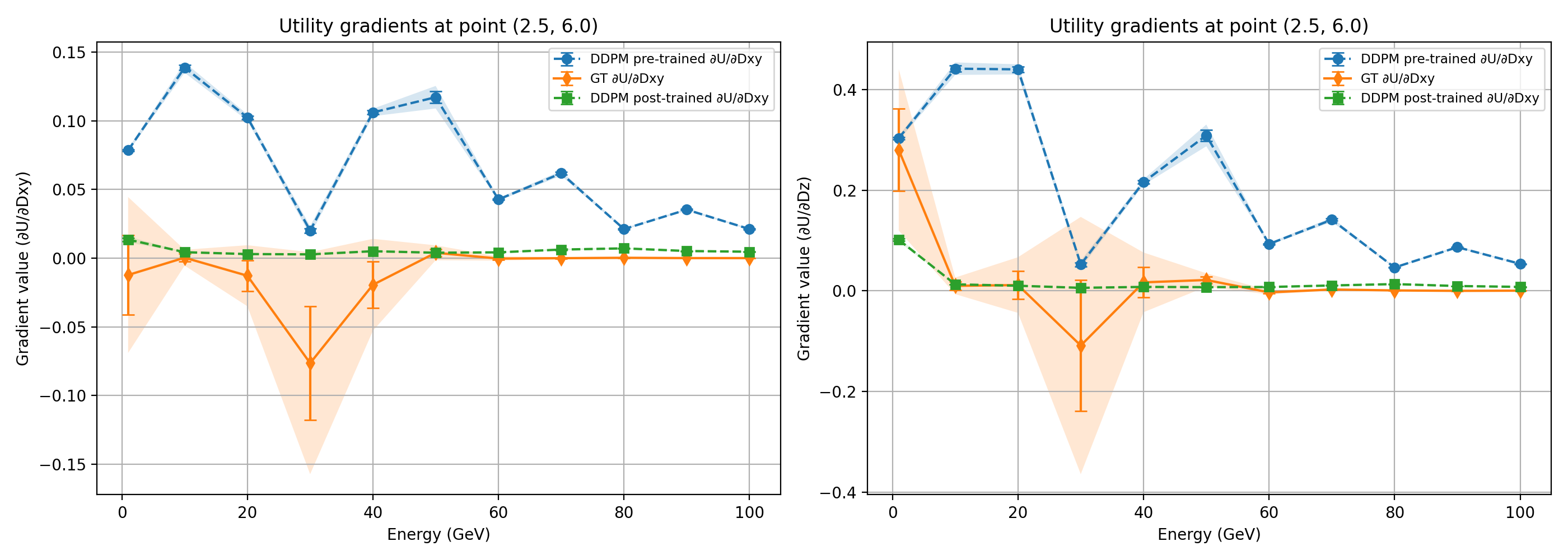}
\caption{Energy dependence of utility gradients at $(\Delta xy, \Delta z)=(2.5,6.0)$\,cm. 
Solid markers denote the mean gradient values and vertical bars indicate $\pm$1 SE. 
Shaded bands show the 95\% confidence interval (mean $\pm 1.96\,$SE). 
Autodiff gradients for the DDPM surrogate were estimated by Monte Carlo averaging over $M$ independent sampling runs; FD gradients were computed from \textsc{GEANT4} samples using central differences and SEs obtained via the delta method applied to per-event MSE values. In some cases, the standard errors are smaller than the marker size and are therefore not visible.}
\label{fig:gradients_prepost}
\end{figure}

The differentiability of the conditional diffusion surrogate was validated by comparing its gradients of the utility function $U$ with respect to the detector design parameters $\Delta xy$ and $\Delta z$ against central FD estimates obtained from \textsc{GEANT4} simulations with a $0.1\,\mathrm{cm}$ step size. To assess gradient stability across energy, Fig.~\ref{fig:gradients_prepost} compares pre-trained and post-trained DDPM gradients at the training configuration point $(\Delta xy,\allowbreak\ \Delta z)$ = (2.5, 6) cm with the finite difference reference. 
Each curve includes uncertainty bars representing $\pm1$ standard error (SE) and shaded bands corresponding to 95\% confidence intervals estimated from independent Monte Carlo runs of the DDPM sampler and 50 FD evaluations. 
The pre-trained model exhibits oscillatory and exaggerated gradients at low energy, whereas post-training substantially reduces this behavior and aligns the gradient magnitude more closely with the reference across the entire 1–100 GeV range. This demonstrates that post-training enhances the surrogate’s local differentiability and improves its quantitative consistency near the training configuration.

Although the absolute magnitudes differ, the DDPM captures the correct gradient sign and geometric dependence within the reported uncertainties. 
While these one-dimensional comparisons demonstrate that the surrogate reconstructs the qualitative behavior of each gradient component, they do not guarantee that the full two-dimensional gradient vector points in the correct direction. 
To assess gradient accuracy more stringently, we therefore complement the component-wise analysis with a vector-level comparison using the cosine similarity between the DDPM and FD gradients.

\begin{figure}[H]
\centering
\includegraphics[width=\textwidth]{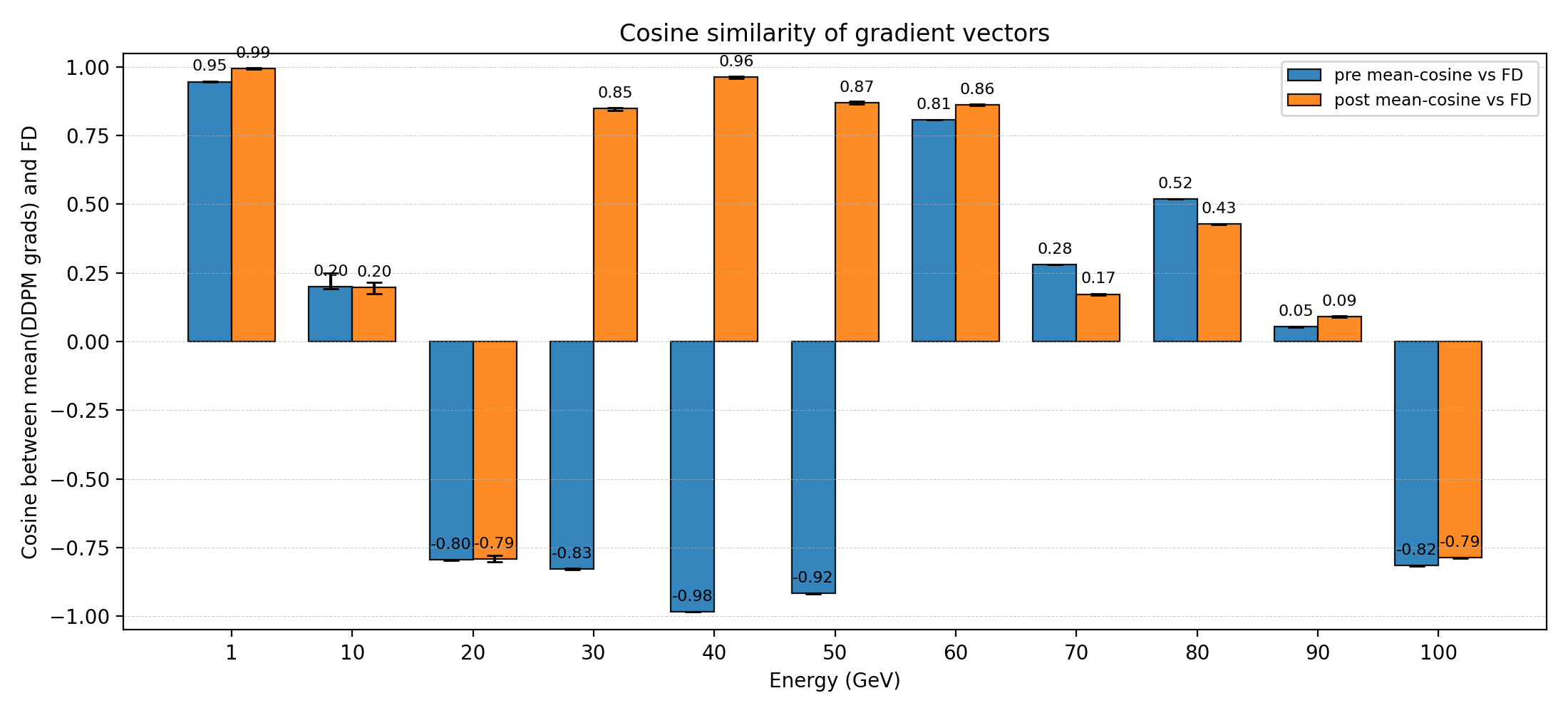}
\caption{Cosine similarity between the mean utility-gradient vectors predicted by the pre-trained and post-trained DDPM surrogates and the FD reference.}
\label{fig:cosine_grad}
\end{figure}

Beyond component-wise comparisons, we also assess the directional consistency of the surrogate’s gradients using cosine similarity between the mean gradient vectors of the DDPM and the FD reference. This metric evaluates whether the surrogate predicts gradients that point in the same direction as the true utility gradient, independent of magnitude rescaling. As shown in Fig.~\ref{fig:cosine_grad}, we compare the cosine similarity between the DDPM-derived utility-gradient vectors and those obtained via FD calculations. While the gradient-alignment fine-tuning improves the similarity at several energies, the agreement remains inconsistent at low and high energies, where sign flips still occur. This highlights that, while the surrogate begins to capture qualitative geometric trends, its gradient predictions remain preliminary.

\section{Conclusion and Outlook}

In this work, we have presented a conditional denoising-diffusion surrogate for electromagnetic calorimeter showers that is both high-fidelity and differentiable with respect to detector-design parameters. The surrogate, built on a U-Net backbone augmented with diffusion-time and calorimeter-specific embeddings, employs DDIM sampling for accelerated inference and gradient-friendly deterministic sampling. Pre-training on a diverse \textsc{GEANT4} corpus yields strong agreement with ground truth on physically meaningful metrics (total energy, energy-weighted radius, and shower dispersion), with RRMSE below 2$\%$ for representative high-energy cases, in line with or exceeding the fidelity of state-of-the-art surrogate approaches such as normalizing-flow and diffusion-based calorimeter models. We further demonstrate that the pre-trained model can be efficiently adapted to an unseen geometry via LoRA using a small post-training set; profile comparisons show substantial improvement after adaptation.

Beyond generative fidelity, we demonstrate that the surrogate’s differentiability enables the computation of physically meaningful utility gradients with respect to detector geometry. Model-derived sensitivities show qualitative agreement with FD references, correctly capturing gradient signs and overall trends across energies and cell sizes. The gradients are smoother and slightly underestimated compared to the ground truth, which can be attributed to the deterministic DDIM sampling process. By removing stochastic variation from the diffusion trajectory, DDIM produces a smooth, single-valued mapping from design parameters to predicted showers, effectively averaging out the local fluctuations present in the underlying physics simulations. However, post-training adaptation improves both magnitude consistency, directional accuracy (as confirmed by cosine similarity), and overall gradient stability. These results establish the surrogate as a viable, differentiable simulator capable of supporting gradient-based calorimeter design optimization.

The present study has several limitations: a restricted set of materials and cell-size combinations, 2D down-sampling of 3D showers for training, the fact that background and detector noise are handled externally rather than being modeled within the generative pipeline and the use of an idealized homogeneous calorimeter that does not explicitly model cell boundary effects, which would need to be included for realistic detector geometries. In future work, we will present the integration of the surrogate into a more sophisticated end-to-end optimization loop, focusing on identifying the optimal configuration of the electromagnetic calorimeter for a muon collider detector. In future work, we plan to improve on these limitations by:
\begin{itemize}
\item extending the training corpus to include more geometries, calorimeter materials, and hadronic showers,
\item incorporating stochastic backgrounds and realistic detector effects,
\item quantifying uncertainty and calibration of surrogate gradients.
\end{itemize}

\section{Acknowledgments}
The authors would like to thank members of the MODE collaboration for valuable discussions that helped
pave the way to this research project.

The authors gratefully acknowledge funding from the German National
High Performance Computing (NHR) association for the
Center NHR South-West.

Pietro Vischia’s work was supported by the "Ramón y Cajal" program under the Project No. RYC2021-033305-I funded by MCIN/AEI/\allowbreak 10.13039/\allowbreak 501100011033 and by the European Union NextGenerationEU/PRTR.

Computing resources have been
provided by the Alliance for High Performance Computing in Rhineland-Palatinate (AHRP) via the
Elwetritsch cluster at the RPTU University Kaiserslautern-Landau, as well as the Artemisa computing infrastructure, funded by the European Union ERDF and the Comunitat Valenciana. The authors also acknowledge the technical support provided by the Instituto de Física Corpuscular (IFIC, CSIC--UV).

\section{Data Availability Statement}
\label{subsec:data_availability}
The source code supporting this work is publicly available at \url{https://github.com/X-T-Nguyen/Diffusion-Surrogate-Detector-Design}.

The datasets used for training and evaluation are available at \url{https://doi.org/10.5281/zenodo.17105137}.

\appendix

\section{Evaluation metrics: definitions and implementation details}
\label{app:metrics_impl}

This appendix provides the full metric definitions, the normalization choices used for RRMSE in the main text, and additional implementation notes for reproducibility.

\subsection{Notation and coordinates}
We denote pixel indices by $(i,j)$ with $i=1\ldots N_x$, $j=1\ldots N_y$, and pixel energies by $I_{ij}$. The physical coordinate of the center of pixel $(i,j)$ is $(x_i,y_j)$; conversions from index to physical coordinate depend on the cell-size family and are applied when reporting metrics in physical units (cm).

\subsection{Full metric definitions}

\begin{itemize}
  \item Total energy: The total deposited energy inside the calorimeter is the sum over all calorimeter cells:
  \begin{equation}\label{app:total_e}
    E \;=\; \sum_{x,y} I(x,y).
  \end{equation}

  \item Energy-weighted centroid: The energy-weighted centroids in \(x\) and \(y\) are
  \begin{equation}\label{app:bar_xy}
    \bar{x} \;=\; \frac{\sum_{x,y} x\,I(x,y)}{\sum_{x,y} I(x,y)}
    \qquad\text{and}\qquad
    \bar{y} \;=\; \frac{\sum_{x,y} y\,I(x,y)}{\sum_{x,y} I(x,y)}.
  \end{equation}

  \item Energy-weighted radius (RMS): The energy-weighted root-mean-square radius \(R_E\) measures the radial spread of the shower:
  \begin{equation}\label{app:R_E}
    R_{E} \;=\; \sqrt{\frac{\sum_{x,y} I(x,y)\big[(x-\bar{x})^2 + (y-\bar{y})^2\big]}{\sum_{x,y} I(x,y)}}.
  \end{equation}

  \item 1D profiles and shower dispersion along \(y\): Define the transverse 1D profiles
  \begin{equation}\label{app:Ix_Iy}
    I_x(x) \;=\; \sum_{y} I(x,y), \qquad
    I_y(y) \;=\; \sum_{x} I(x,y).
  \end{equation}
  The energy-weighted \(y\)-centroid computed from the profile \(I_y\) is
  \begin{equation}\label{app:y_bar}
    \bar{y} \;=\; \frac{\sum_{y} y\,I_{y}(y)}{\sum_{y} I_{y}(y)}.
  \end{equation}
  The dispersion (standard deviation) along \(y\) is then
  \begin{equation}\label{app:sigma_y}
    \sigma_{y} \;=\; \frac{\sum_{y} (y-\bar{y})^2\,I_{y}(y)}{\sum_{y} I_{y}(y)}.
  \end{equation}
\end{itemize}

\subsection{Dataset-level comparison and normalization}
Given $N$ events, for a generic per-event metric $M$ we compute:
\begin{equation}\label{app:eq:MSE}
  \text{MSE}_{M} \;=\; \frac{1}{N}\sum_{i=1}^N \big(M^{(i)}_{\text{gen}} - M^{(i)}_{\text{GT}}\big)^2 .
\end{equation}

The quantity reported in the main text is the \emph{relative} RMSE:
\begin{equation}\label{app:eq:RRMSE}
  \text{RRMSE}_{M} \;=\; \frac{\sqrt{\text{MSE}_{M}}}{\overline{M_{\text{GT}}}}
  \quad\text{where}\quad
  \overline{M_{\text{GT}}}=\frac{1}{N}\sum_{i=1}^N M^{(i)}_{\text{GT}}.
\end{equation}

For the total-energy metric we use the notation \(E_{\text{true}}\) to denote the event's true total deposited energy (ground truth); when discussing total-energy errors in the text we normalize consistently using the convention in Eq.~\eqref{app:eq:RRMSE} with $\overline{M_{\text{GT}}}$ or, when reporting per-energy-bin relative errors, by the corresponding $E_{\text{true}}$ as appropriate. 

\subsection{Implementation notes (reproducibility)}
\begin{itemize}
  \item \textbf{Coordinates:} $x,y$ are pixel indices (0-based) on the $32\times32$ down-sampled grid. When comparing across cell-size families we convert pixel indices to physical coordinates using the per-configuration pixel size; this conversion is applied when computing $R_E$ and $\sigma_y$ reported in physical units (cm).
  \item \textbf{Binning:} when projecting 3D→2D we sum energy in depth layers onto the transverse grid. The exact projection (e.g., sum vs peak slice) used for each figure is listed in the code repository referenced in the supplement.
  \item \textbf{Edge cases:} events with zero total deposited energy (rare in our \textsc{GEANT4} samples) are excluded from mean computations; the number of excluded events is reported in the captions where relevant.

\end{itemize}

\section{Generated Energy Depositions}
\begin{figure}[H] 
    \centering

    \includegraphics[width=0.95\linewidth]{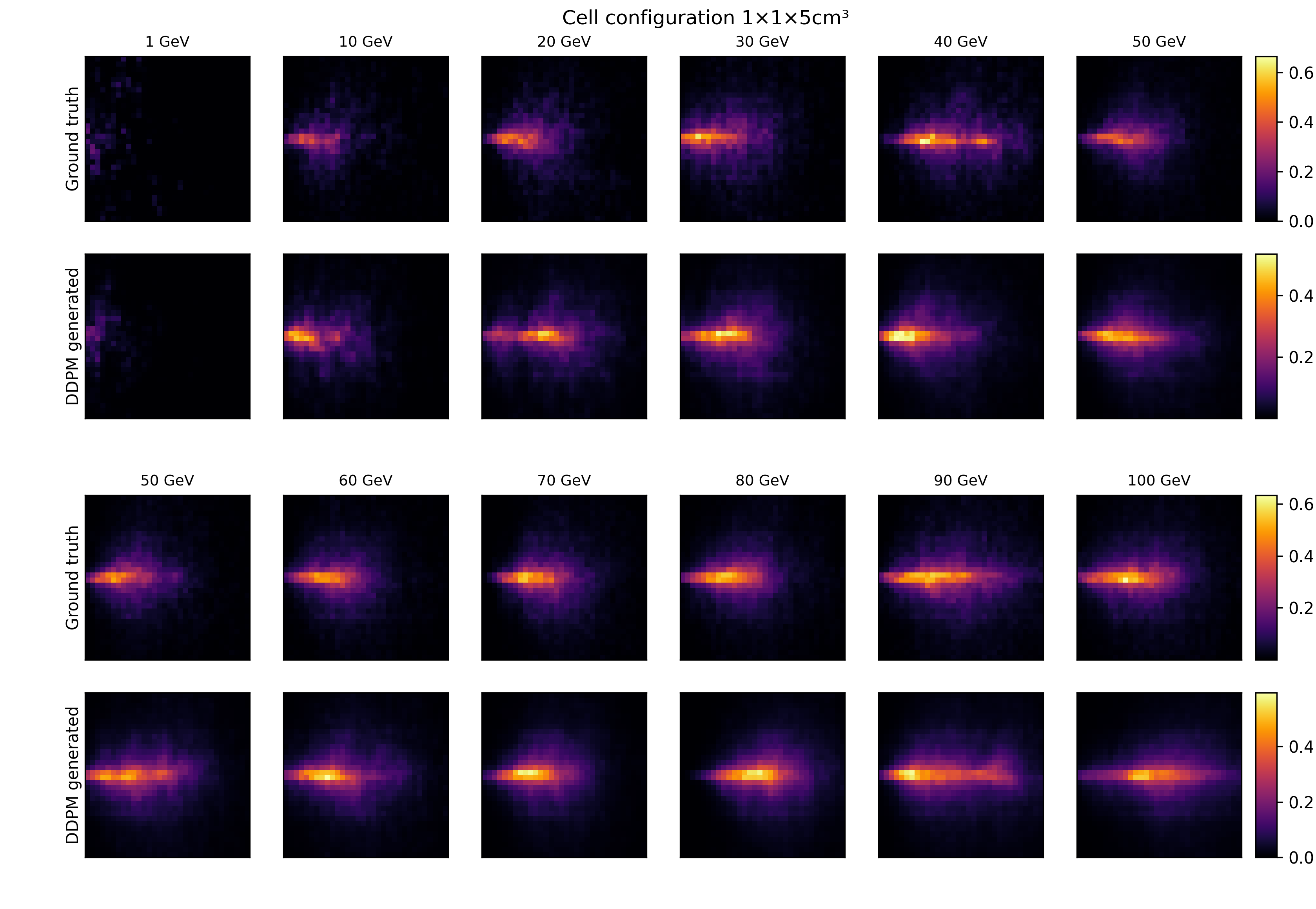}    

    \includegraphics[width=0.95\linewidth]{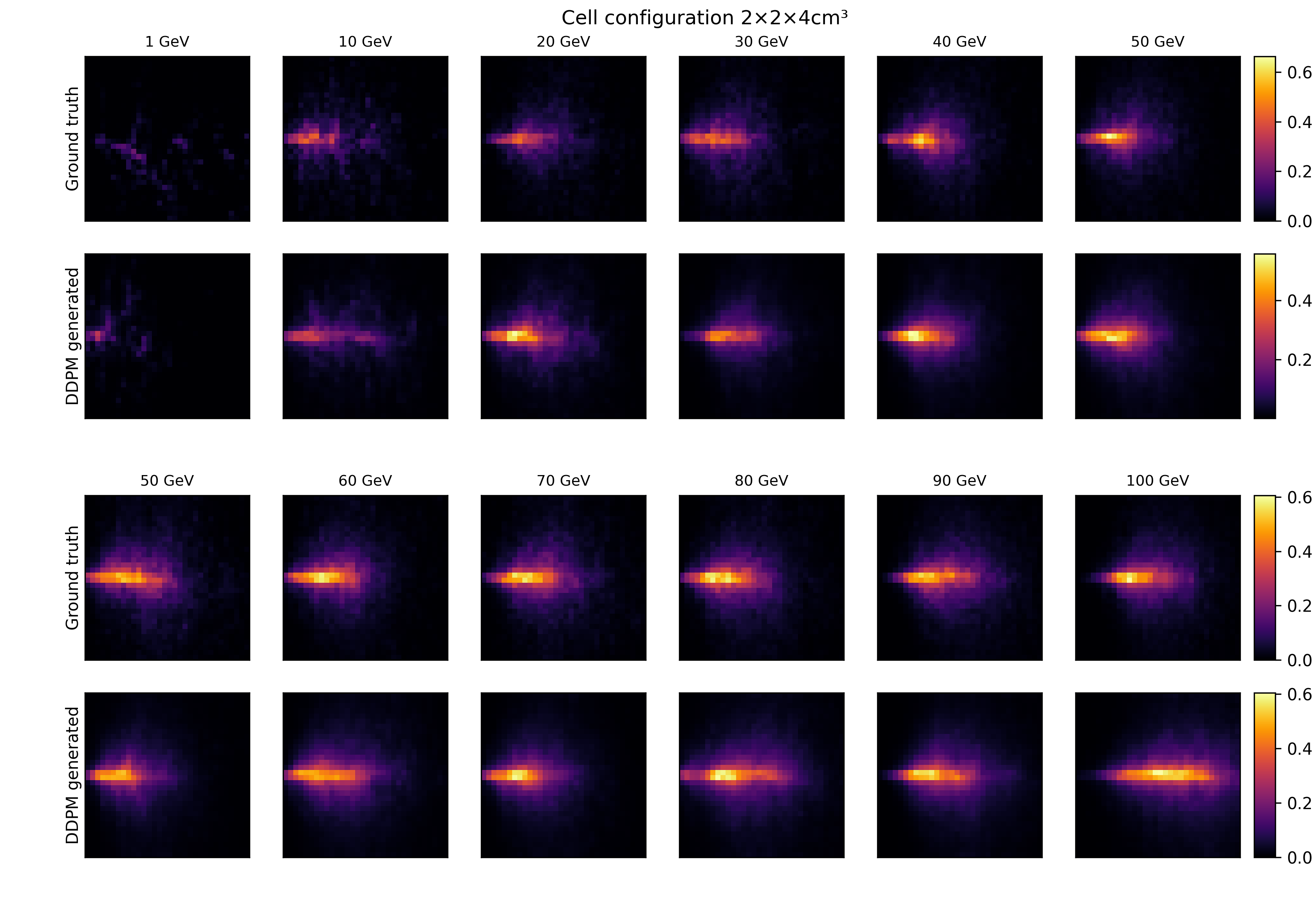}
    
    \caption{Comparison of Monte Carlo and diffusion-generated energy-deposition maps for photon showers at 1, 10, 20, 30, 40, 50, 60, 70, 80, 90, and 100 GeV, shown for two calorimeter geometries. The top two pairs of rows correspond to $1\times1\times5 \mathrm{cm}^3$ the bottom two pairs correspond to $2\times2\times4 \mathrm{cm}^3$ cells.}

    \label{fig:compare_showers_all_1}
\end{figure}

\begin{figure}[H] 
    \centering

    \includegraphics[width=0.95\linewidth]{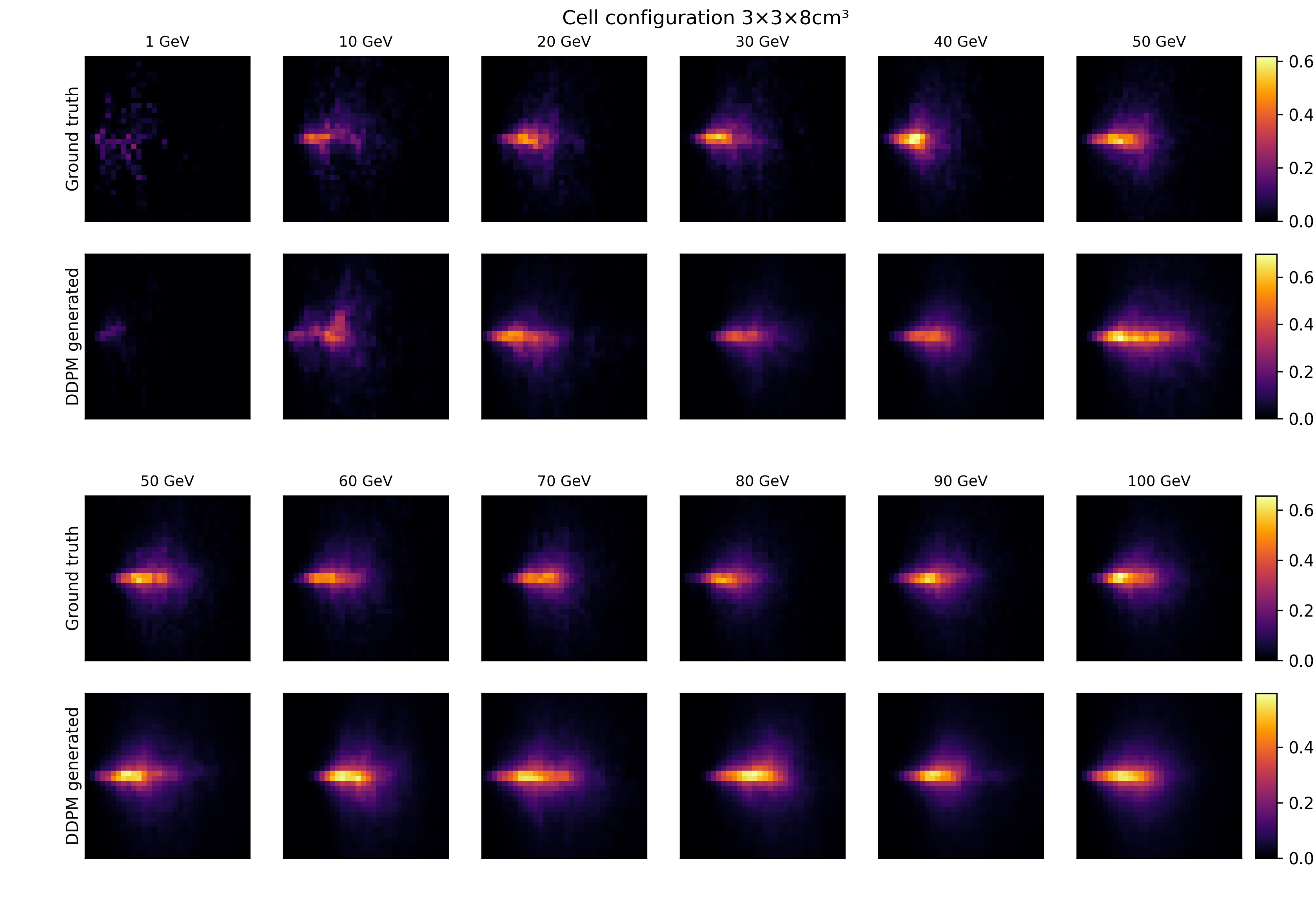}    

    \includegraphics[width=0.95\linewidth]{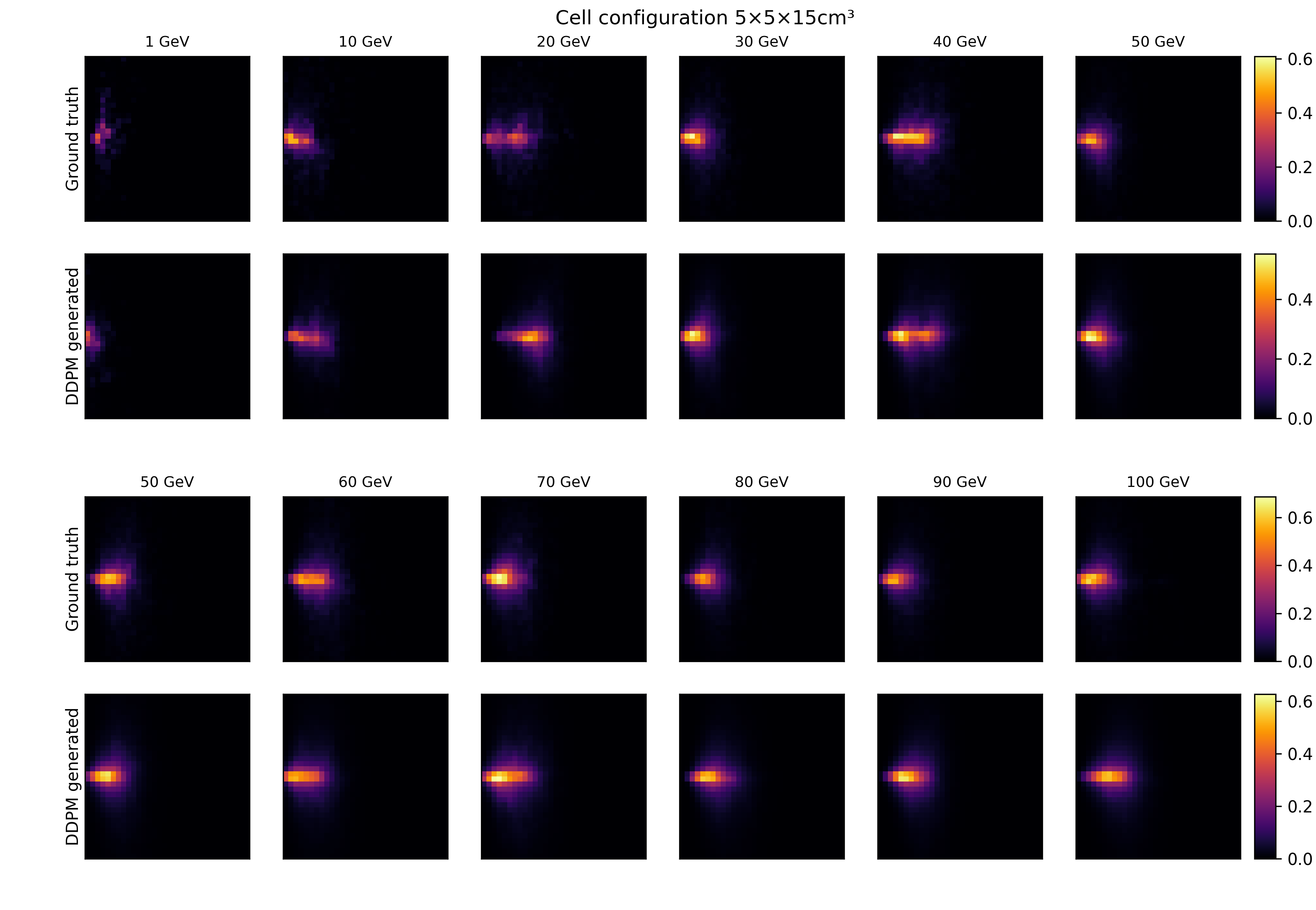}
    
    \caption{Comparison of Monte Carlo and diffusion-generated energy-deposition maps for photon showers at 1, 10, 20, 30, 40, 50, 60, 70, 80, 90, and 100 GeV, shown for two calorimeter geometries. The top two pairs of rows correspond to $3\times3\times8 \mathrm{cm}^3$ the bottom two pairs correspond to $5\times5\times15 \mathrm{cm}^3$ cells.}
    \label{fig:compare_showers_all_2}
\end{figure}

Figures~\ref{fig:compare_showers_all_1} and~\ref{fig:compare_showers_all_2} show 2D projections of calorimeter showers generated by the DDPM model for the following cell configurations $1\times1\times5$, $2\times2\times4$, $3\times3\times8$, and $5\times5\times15\ \mathrm{cm}^3$. For each energy and geometry, the top row displays ground-truth histograms from the down-sampled ($32\times32$) dataset, while the bottom row shows samples generated by the conditional DDPM under the same normalized (xy, z, material, energy) conditions.
All panels use the \texttt{inferno} colormap from Matplotlib; columns are labeled by the primary photon energy.

\section{Energy Profiles}
\begin{figure}[h]
    \centering
    \includegraphics[width=0.95\linewidth]{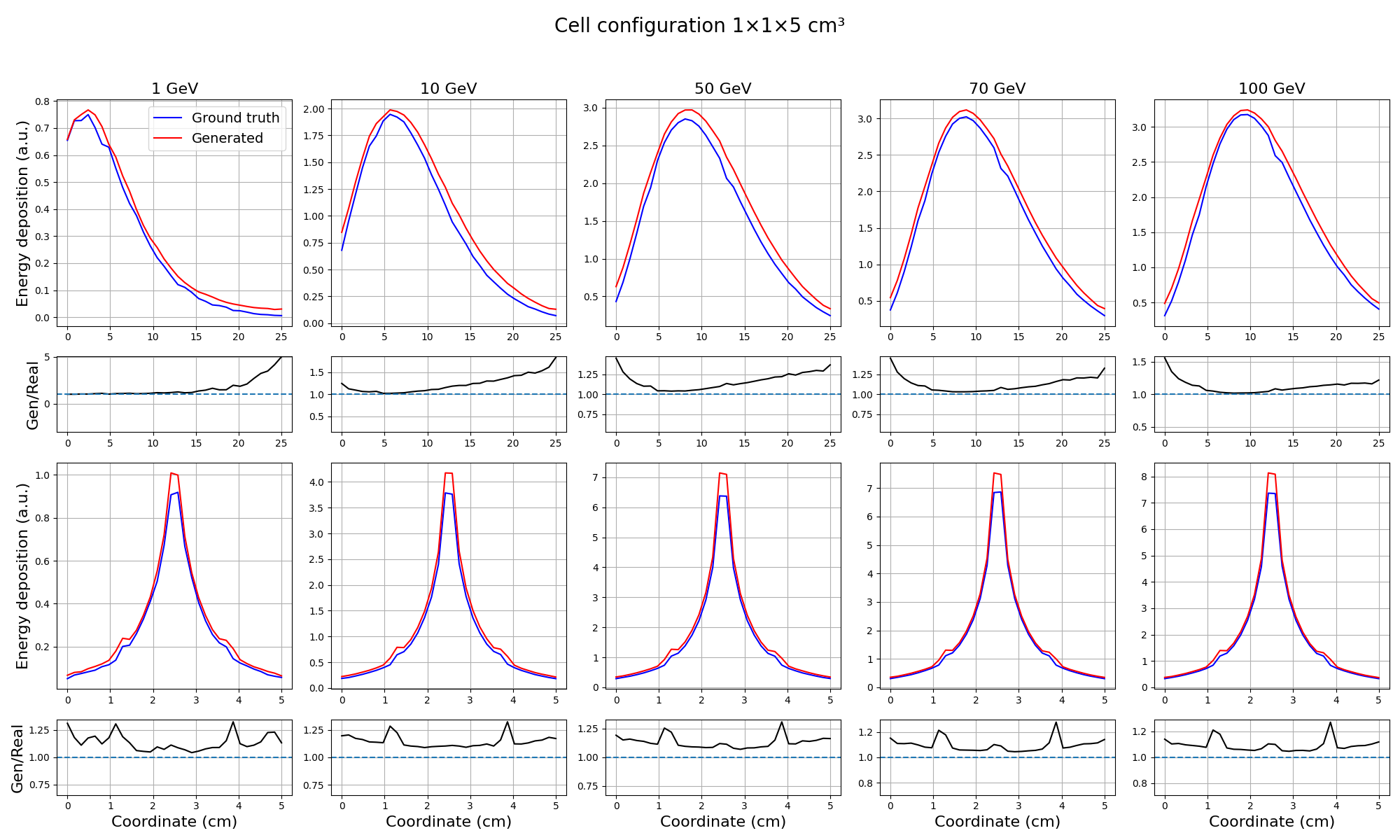}
    \includegraphics[width=0.95\linewidth]{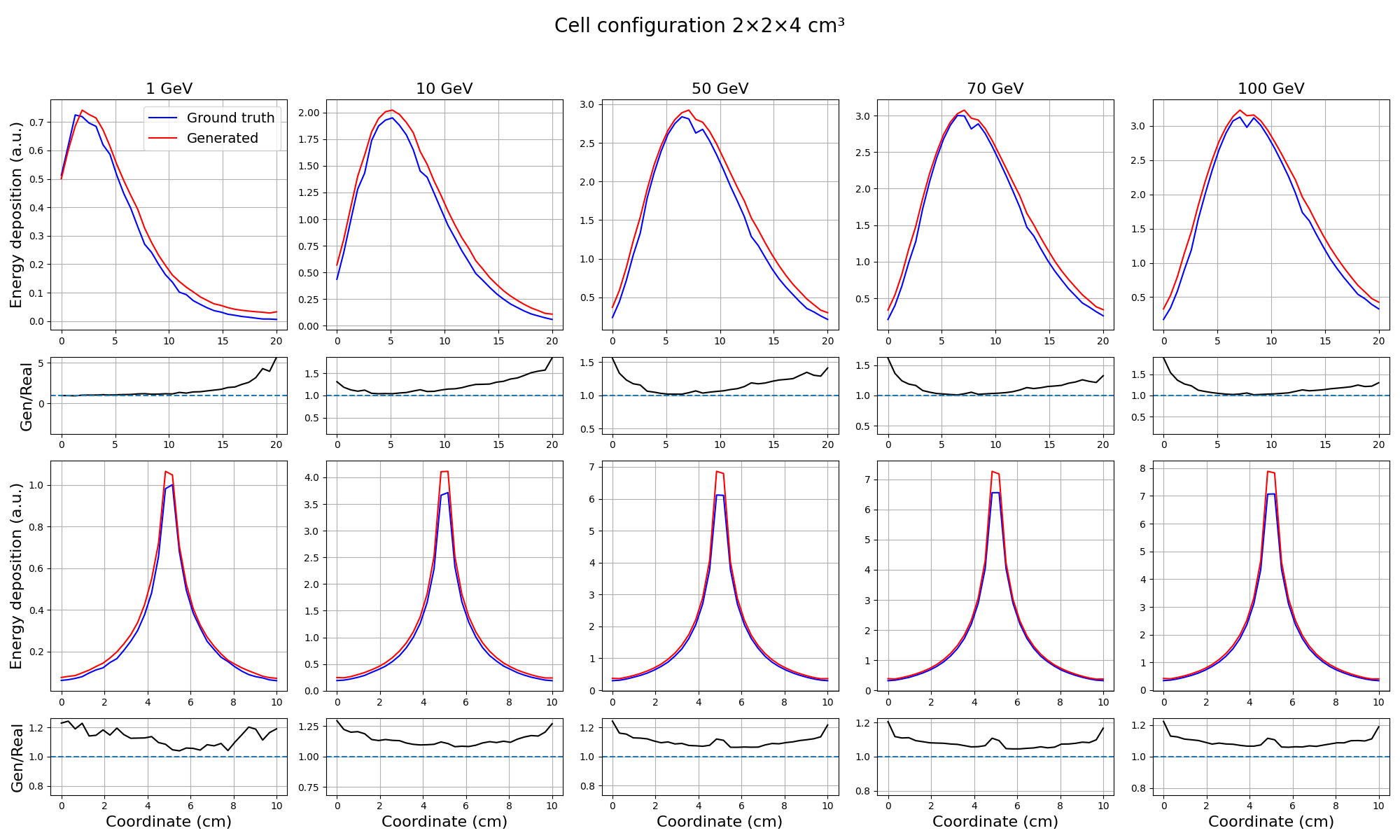}

    \caption{Longitudinal (top row) and transverse (bottom row) average energy deposition profiles across primary energies from 1 to 100 GeV. Blue curves denote real samples from the down-sampled \textsc{GEANT4} dataset ($32 \times 32$), while red curves indicate generated samples from our conditional DDPM model under the same (normalized) conditions. The longitudinal profile is averaged over transverse coordinates ($x$ and $y$), and the transverse profile is taken at the shower maximum in depth. Each subplot corresponds to a specific primary energy.}
    \label{fig:profiles-all_1}
\end{figure}

\begin{figure}[h]
    \centering
    \includegraphics[width=0.95\linewidth]{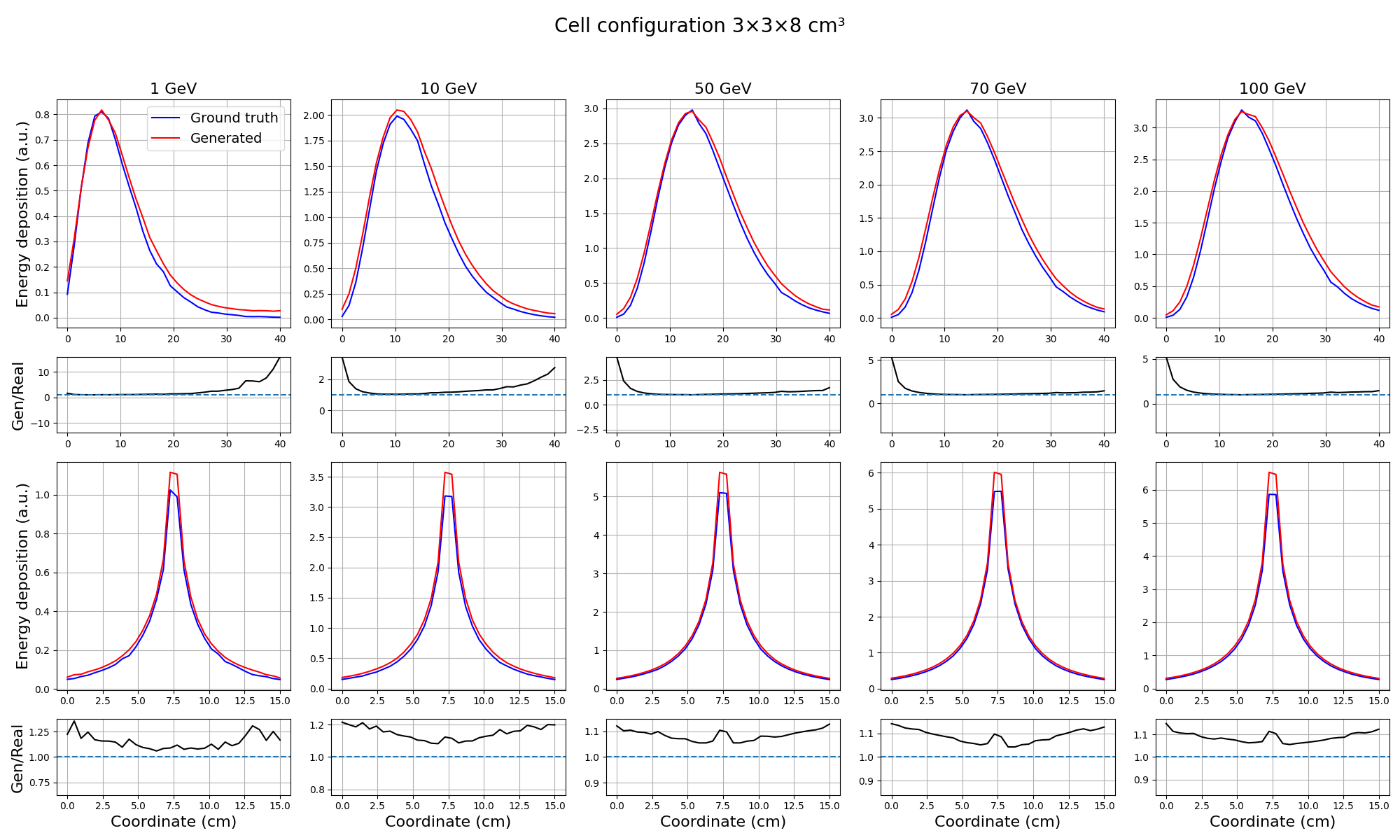}
    \includegraphics[width=0.95\linewidth]{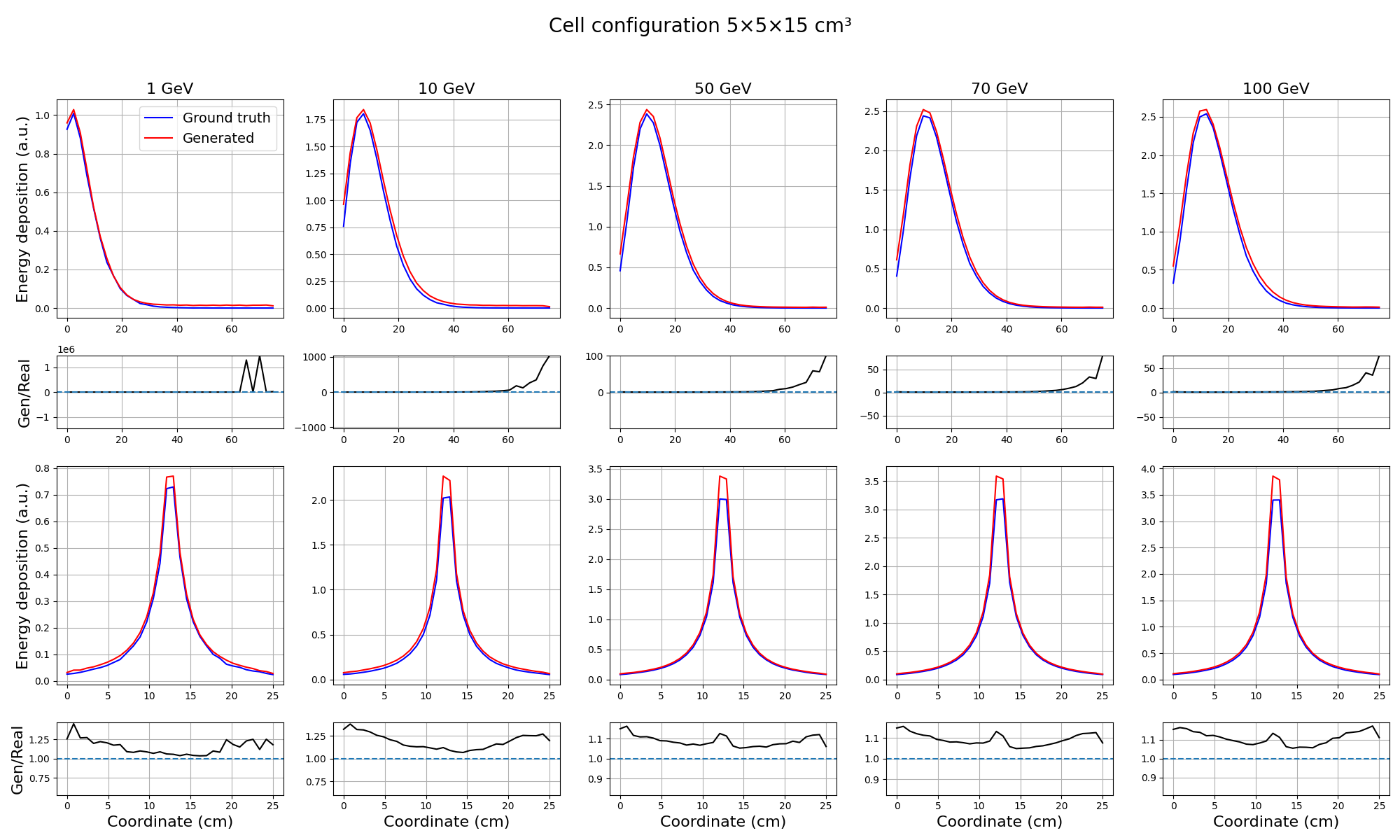}

    \caption{Longitudinal (top row) and transverse (bottom row) average energy deposition profiles across primary energies from 1 to 100 GeV. Blue curves denote real samples from the down-sampled \textsc{GEANT4} dataset ($32 \times 32$), while red curves indicate generated samples from our conditional DDPM model under the same (normalized) conditions. The longitudinal profile is averaged over transverse coordinates ($x$ and $y$), and the transverse profile is taken at the shower maximum in depth. Each subplot corresponds to a specific primary energy.}
    \label{fig:profiles-all_2}
\end{figure}

Figures~\ref{fig:profiles-all_1} and \ref{fig:profiles-all_2} show the average longitudinal energy profiles produced by the model compared to ground-truth for each configuration and energy level.

\begin{figure}[h]
    \centering
    \includegraphics[width=0.95\linewidth]{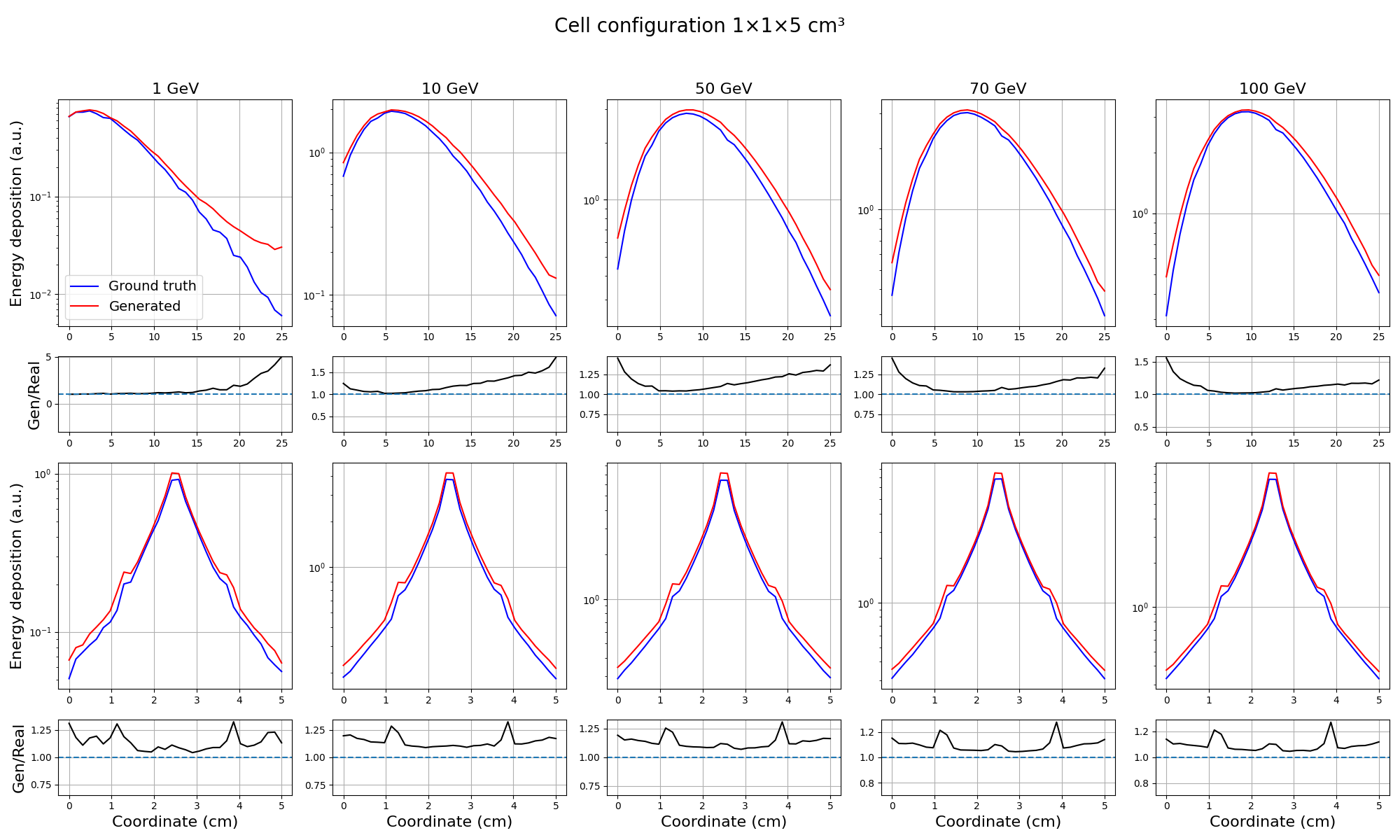}
    \includegraphics[width=0.95\linewidth]{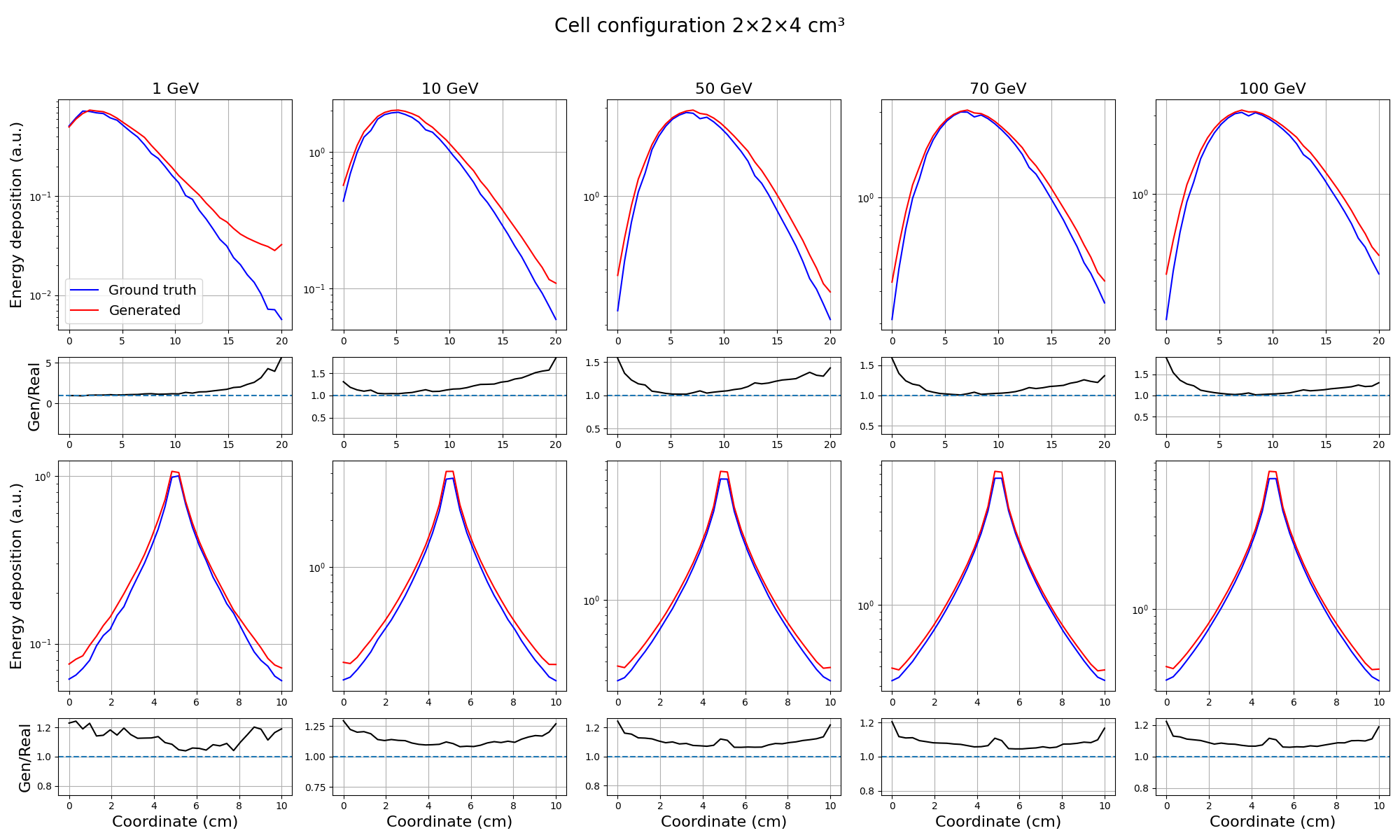}

    \caption{Longitudinal (top row) and transverse (bottom row) average energy deposition profiles on a logarithmic scale for the same configuration as Fig.~\ref{fig:profiles-all_1}. The log scale enhances the visibility of low-energy regions, enabling a more detailed comparison between ground-truth (blue) and DDPM-generated (red) distributions across primary energies.}
    \label{fig:profiles-all_log_1}
\end{figure}

\begin{figure}[h]
    \centering
    \includegraphics[width=0.95\linewidth]{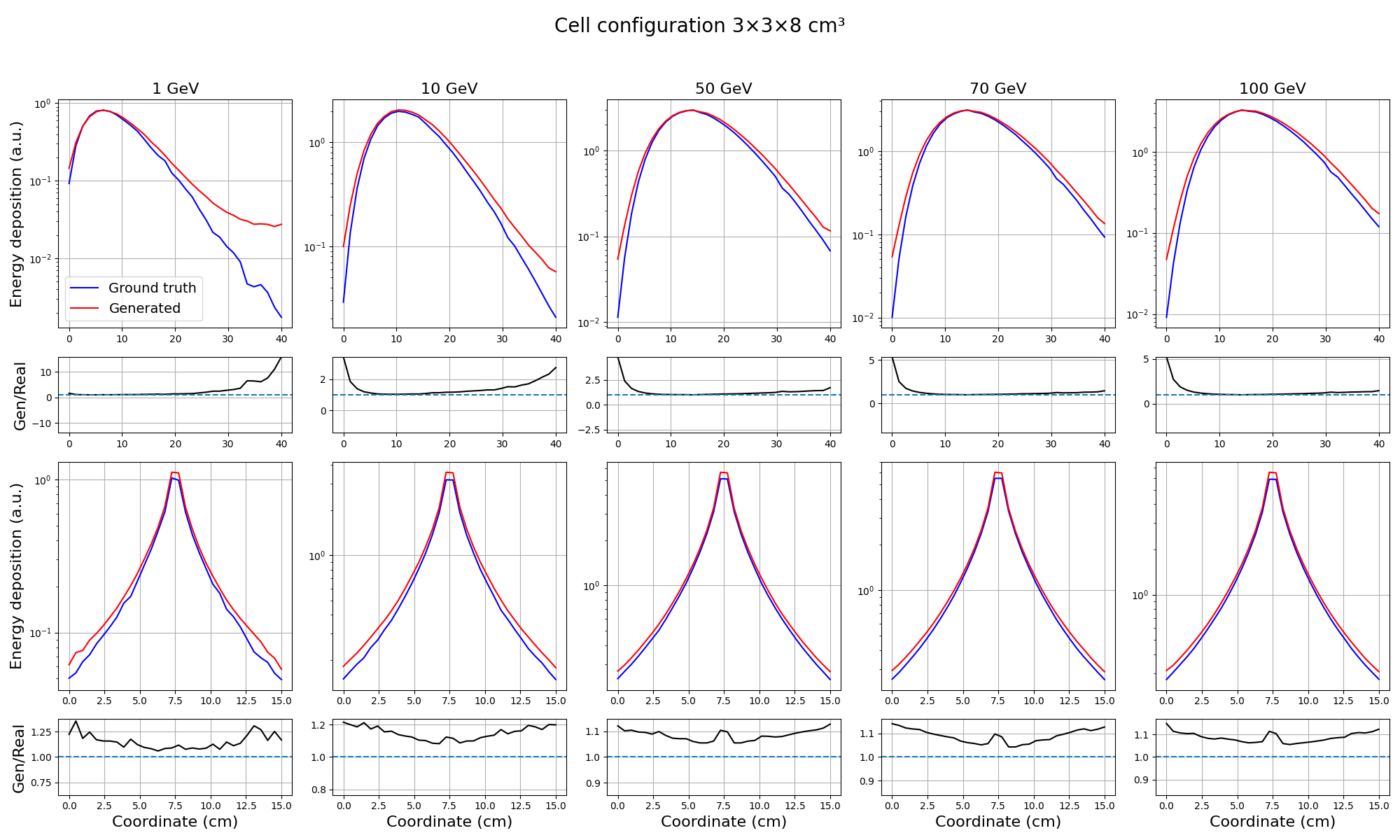}
    \includegraphics[width=0.95\linewidth]{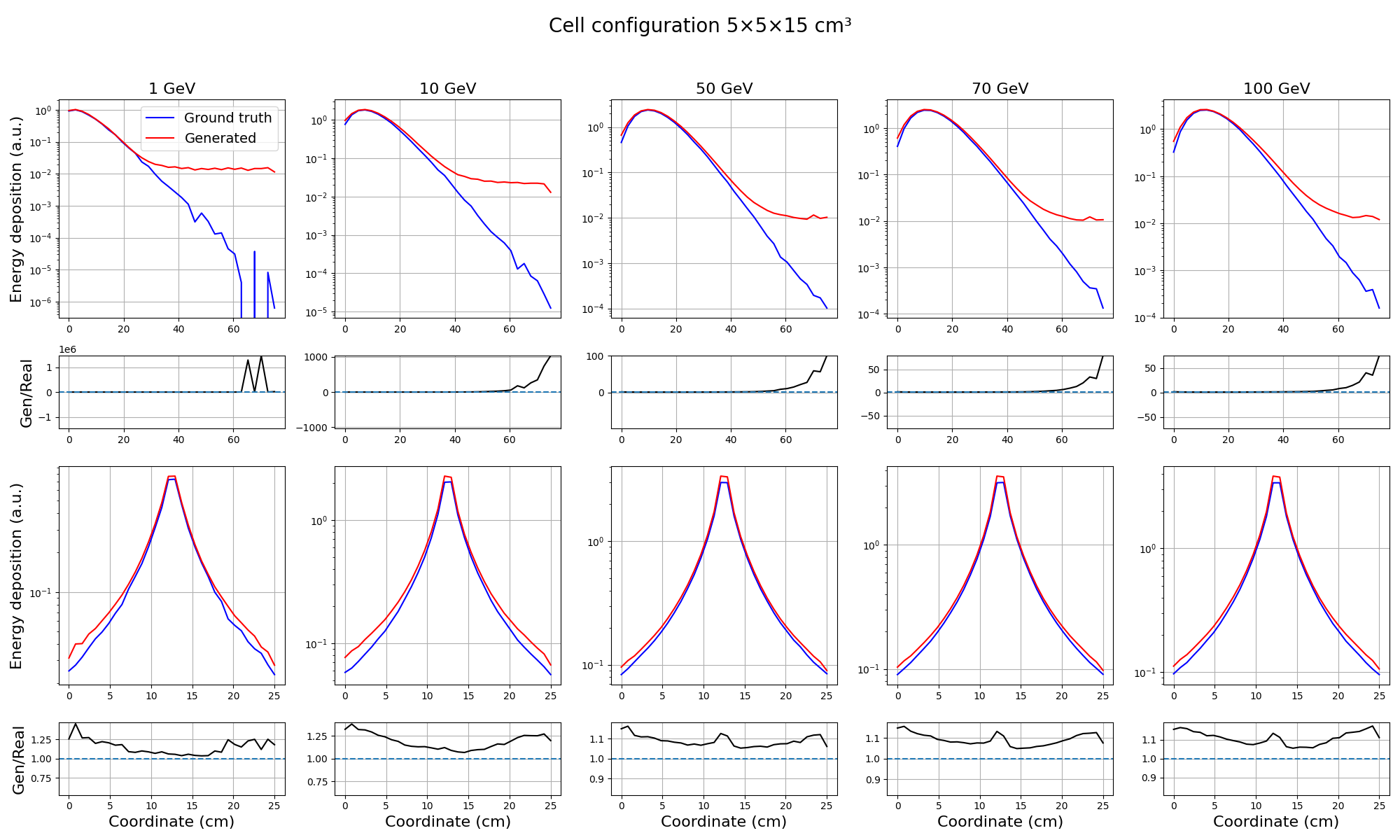}

    \caption{Longitudinal (top row) and transverse (bottom row) average energy deposition profiles on a logarithmic scale for the same configuration as Fig.~\ref{fig:profiles-all_2}. The log scale enhances the visibility of low-energy regions, enabling a more detailed comparison between ground-truth (blue) and DDPM-generated (red) distributions across primary energies.}
    \label{fig:profiles-all_log_2}
\end{figure}

Figures~\ref{fig:profiles-all_log_1} and \ref{fig:profiles-all_log_2} show the same profiles on a logarithmic scale to highlight low-energy regions.

\clearpage

\clearpage

\section{Evaluation Metrics}
\begin{figure}[H]
    \centering
    \includegraphics[height=0.45\textheight]{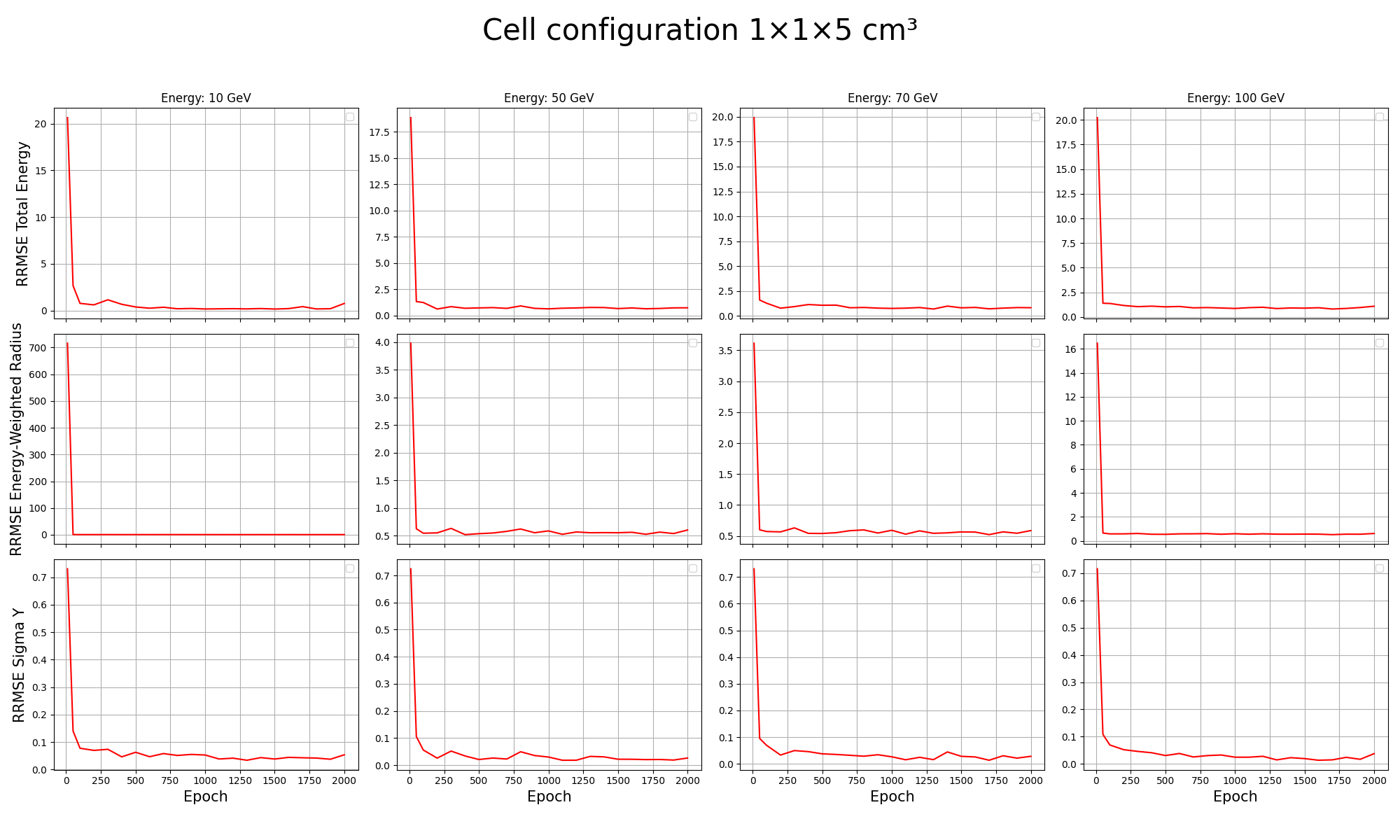}
    
    \includegraphics[height=0.45\textheight]{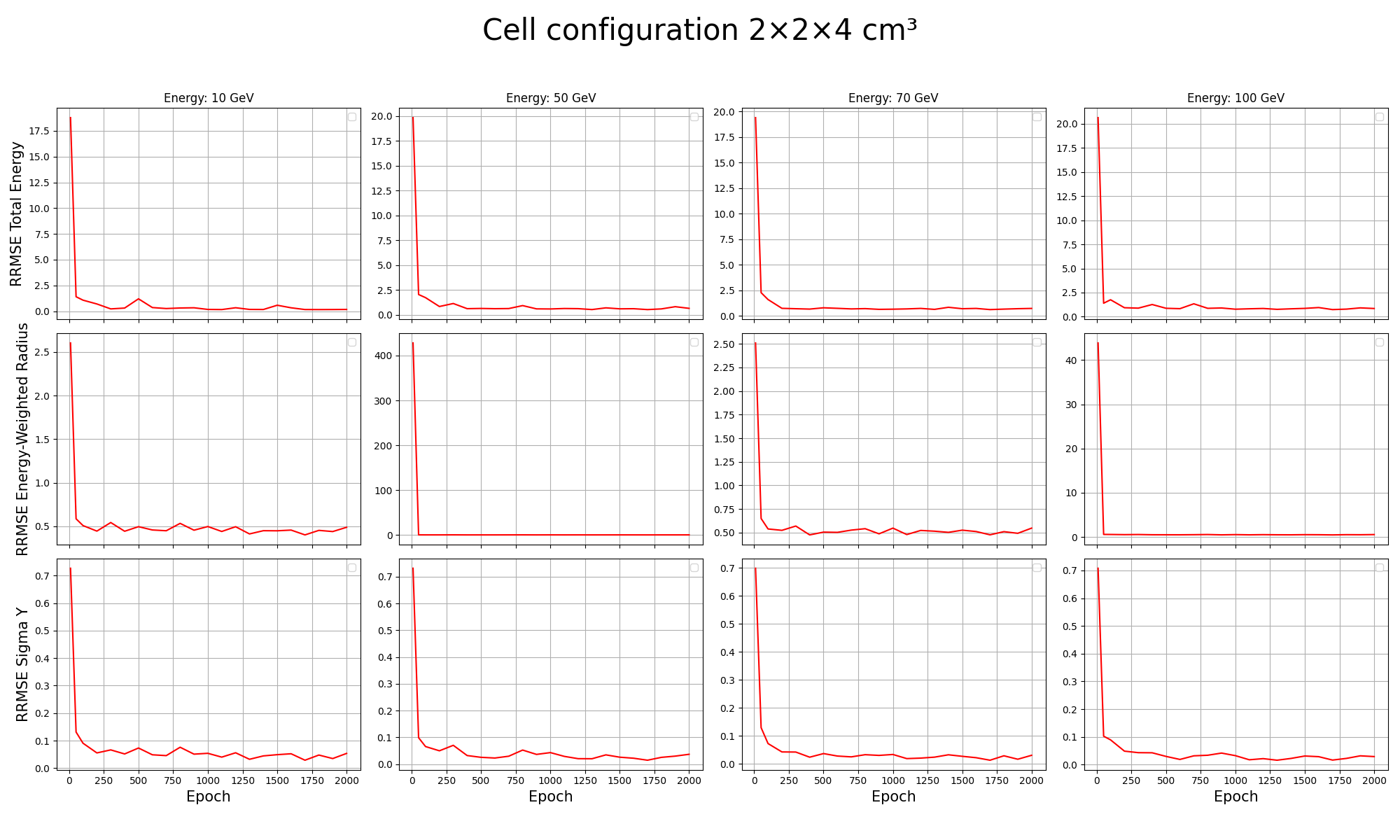}
    
    \caption{RRMSE vs training epoch for different calorimeter configurations. Each plot shows total energy (top), energy-weighted radius (middle), and dispersion $\sigma_y$ (bottom) across primary energies.}
    \label{fig:metrics_all_1}
\end{figure}

\begin{figure}[H]
    \centering
    
    \includegraphics[height=0.45\textheight]{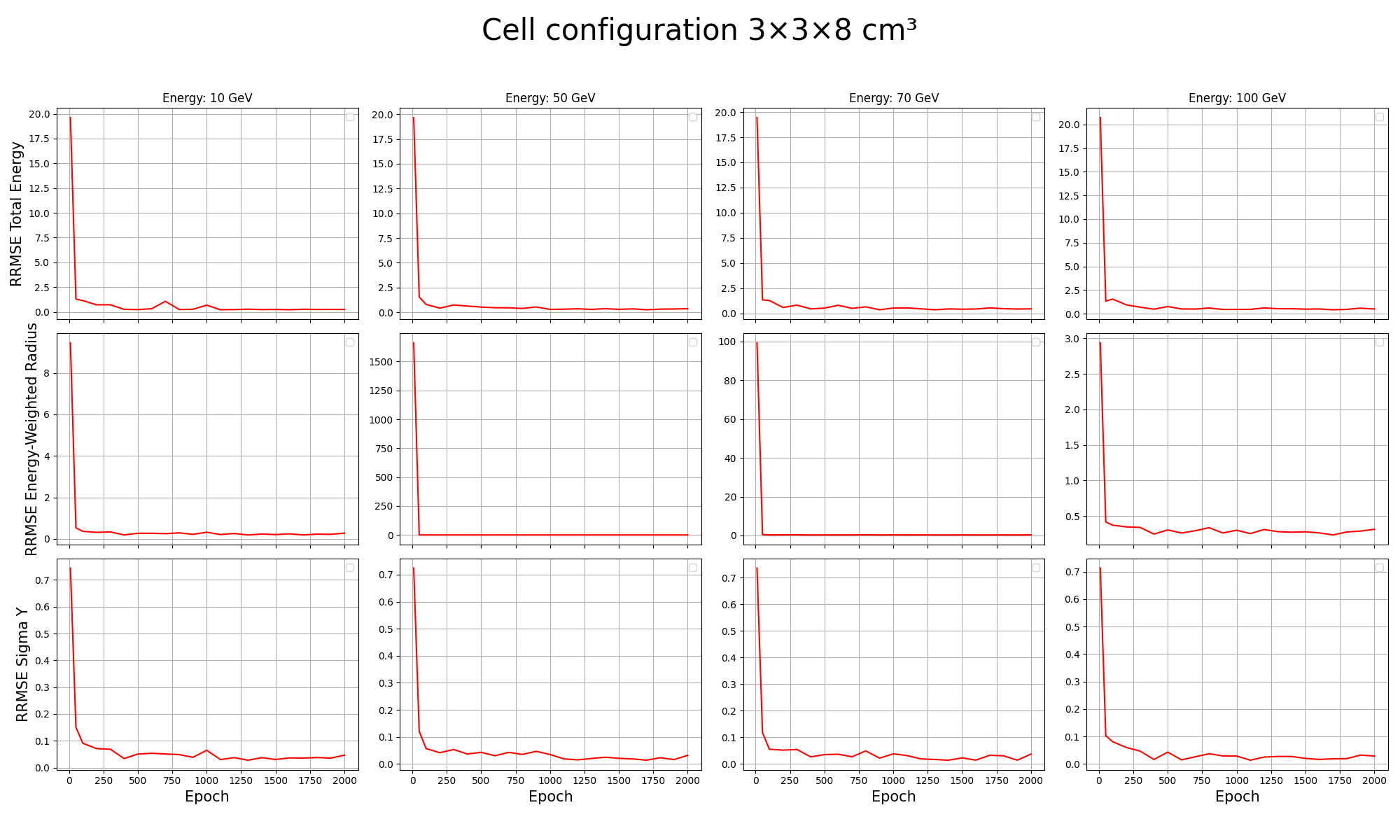}
    
    \includegraphics[height=0.45\textheight]{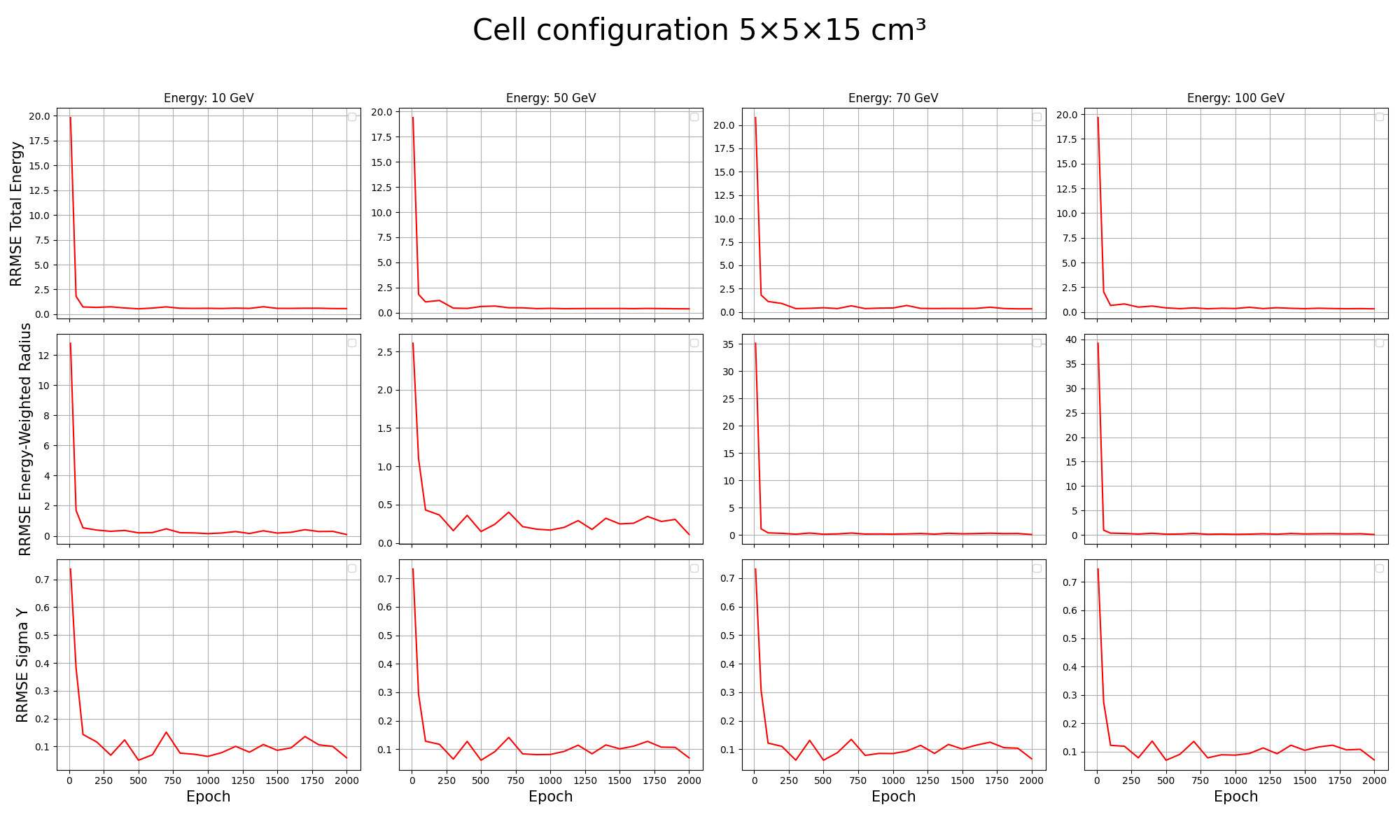}
    
    \caption{RRMSE vs training epoch for different calorimeter configurations. Each plot shows total energy (top), energy-weighted radius (middle), and dispersion $\sigma_y$ (bottom) across primary energies.}
    \label{fig:metrics_all_2}
\end{figure}

Figures~\ref{fig:metrics_all_1} and \ref{fig:metrics_all_2} show the quantitative evaluation (e.g., total energy, radial energy, shower dispersion, RRMSE) for each configuration.

\bibliographystyle{elsarticle-num} 
\bibliography{reference}







\end{document}